\documentclass[fleqn,10pt]{article}
\usepackage[utf8]{inputenc}
\usepackage[T1]{fontenc}
\usepackage{lineno}
\usepackage{adjustbox}
\usepackage{graphicx}
\usepackage{multirow}
\usepackage{geometry}
\usepackage{booktabs}
\usepackage{xcolor} 
\usepackage{times}
\usepackage{tipa}
\usepackage{subfig}
\usepackage{courier}
\usepackage{parskip} 
\usepackage{tipa}
\usepackage{acronym}
\usepackage{soul}
\usepackage{url}

\newcommand{\rw}[1]{{\color{black}{#1}}}
\newcommand{\julian}[1]{{\color{black}{#1}}}

\providecommand\JournalTitle[1]{#1}

\acrodef{ByO}{Bioengineering and Optoelectronics}
\acrodef{CNN}{Convolutional Neural Network}
\acrodef{GRBAS}{Grade, Roughness, Breathiness, Asthenia, and Strain}
\acrodef{DDK}{diadochokinetic}
\acrodef{DL}{Deep Learning}
\acrodef{IPA}{International Phonetic Alphabet}
\acrodef{HC}{Healthy Control}
\acrodef{HGUGM}{Hospital General Universitario Gregorio Marañón}
\acrodef{HUF}{Hospital Universitario de Fuenlabrada}
\acrodef{HNR}{Harmonics-to-Noise Ratio}
\acrodef{IPA}{International Phonetic Alphabet}
\acrodef{H-Y}{Hoehn and Yahr}
\acrodef{LR}{Listen and Repeat}
\acrodef{MDS}{International Parkinson and Movement Disorders Society}
\acrodef{ML}{Machine Learning}
\acrodef{PD}{Parkinson's Disease}
\acrodef{SPL}{Sound Pressure Level}
\acrodef{SNR}{Signal-to-Noise}
\acrodef{UPDRS}{Unified Parkinson's Disease Rating Scale}
\acrodef{UPM}{Universidad Politécnica de Madrid}

\title{NeuroVoz: a Castillian Spanish corpus of parkinsonian speech}

\author{Janaína Mendes-Laureano$^1$ \and Jorge A. Gómez-García$^1$ \and Alejandro Guerrero-López$^1$ \and Elisa Luque-Buzo$^2$ \and Julián D. Arias-Londoño$^1$ \and Francisco J. Grandas-Pérez$^2$ \and Juan I. Godino-Llorente$^1$\thanks{Corresponding author: ignacio.godino@upm.es}}

\date{
	$^1$Escuela Técnica Superior de Ingenieros de Telecomunicación, Universidad Politécnica de Madrid, 28040, Madrid, Spain\\%
	$^2$Department of Neurology, Hospital General Universitario Gregorio Marañón, 28007, Madrid, Spain \\
}

\begin{document}
\maketitle
\begin{abstract}


The screening of Parkinson's Disease (PD) through speech is hindered by a notable lack of publicly available datasets in different languages. This fact limits the reproducibility and further exploration of existing research.

\rw{To address this gap, this manuscript presents the NeuroVoz corpus consisting of} 112 native Castilian-Spanish speakers, including 58 healthy controls and 54 individuals with PD, all recorded in ON state. The corpus showcases a diverse array of speech tasks: sustained vowels; diadochokinetic tests; 16 Listen-and-Repeat utterances; and spontaneous monologues. 

The dataset is also complemented with subjective assessments of voice quality performed by an expert according to the GRBAS scale (Grade/Roughness/Breathiness/Asthenia/Strain), as well as annotations with a thorough examination of phonation quality, intensity, speed, resonance, intelligibility, and prosody. 

The corpus offers a substantial resource for the exploration of the impact of PD on speech. This data set has already supported several studies, achieving a benchmark accuracy of 89\% for the screening of PD. Despite these advances, the broader challenge of conducting a language-agnostic, cross-corpora analysis of Parkinsonian speech patterns remains open. 


\end{abstract}



\thispagestyle{empty}

\noindent 




\section*{Background}




\ac{PD} is a progressive neurodegenerative disorder primarily associated with ageing. \rw{Although it predominantly affects people over 65 year, it is a common and worldwide condition affecting approximately 61 million people in 2016\cite{bloem2021parkinson}.}
\ac{PD} is characterised by neuronal degeneration in the substantia nigra, a region of the brain responsible for the production of dopamine \cite{aarsland2017cognitive}. 
The resulting decrease in dopamine levels leads to various motor symptoms, including speech impairments, coordination difficulties, muscle rigidity, bradykinesia, among others, significantly affecting the quality of life of those affected \cite{friedman2016fatigue}. In addition to motor problems, patients with \ac{PD} often experience non-motor symptoms such as cognitive impairment, mood disturbances, and autonomic dysfunction \cite{pfeiffer2016non}.

The diagnosis of \ac{PD} is primarily based on the evaluation of motor and non-motor symptoms \cite{pfeiffer2016non}, with a final confirmation that can only be provided by postmortem neuropathological examination \cite{koga2015dlb}.
Despite the clinical precision achieved, which is approximately 90\%, reaching a definitive diagnosis can take an \rw{average of 2.75 years\cite{rossi2021much}}. In addition to that, the varied symptomatology of \ac{PD} and its overlap with other neurodegenerative diseases highlight the need for more sophisticated diagnostic tools to improve evaluation procedures \cite{tolosa2021challenges}. 

Recent advances\cite{Pujols2018SmallMI} underscore the potential of the integration of novel biomarkers and \ac{ML} algorithms to improve diagnostic precision and accelerate disease identification, thus facilitating earlier intervention and potentially improving the quality of life of patients with \ac{PD}.
In this regard, speech, with its precise and coordinated movements, holds promise as a potential biomarker to detect and assess \ac{PD}.
Speech impairments are common in approximately 89\% of patients with \ac{PD}\cite{ramig2008speech}, and are believed to be the result of a combination of broad motor deficits and specific difficulties related to the breathing and swallowing processes \cite{robbins1986swallowing}. 
These impairments include deficits in voice production (e.g., dysponia), in articulation (e.g., dysarthria), and in prosody (e.g., disprosody) \cite{weismer1984articulatory, skodda2011vowel, rusz2011quantitative,ackermann1991articulatory, kegl1999articulatory, duffy2014motor, moro2021advances,moro2020review}. 
The literature has documented articulatory deficits in patients with \ac{PD}, including reduced precision and variability in speech organ movements, leading to notable changes in speech patterns \cite{Walsh2012BasicPO, Svensson1993SpeechMC}. 
These include variations in voice onset time\cite{Chenausky2011AcousticAO}, consonant spirantization\cite{GodinoLlorente2017TowardsTI, Antolk2013ConsonantDI}, altered formant frequencies \cite{Weismer1998FormantTC}, reduced area of the vowel space\cite{Rusz2013ImpreciseVA}, and changes in speech rate \cite{McRae2002AcousticAP}, which collectively serve as potential indicators of the presence of \ac{PD}.
Regarding phonation deficits,
\ac{PD} patients' voice has been identified with hoarseness, tremor, roughness \cite{logemann1978frequency} or with breathiness and asthenia\cite{midi2008voice}.
There is also evidence of higher levels of noise and irregularity due to incomplete closure of the vocal folds, phase asymmetry, or vocal tremor, which have been verified instrumentally \cite{perez1996parkinson} and through noise measurements (jitter and \ac{HNR}) \cite{ramig1988acoustic, rahn2007phonatory, tanaka2011vocal, rusz2013evaluation}. 
Similarly, variability in fundamental frequency has been identified to be significantly different in \ac{PD} vs. controls \cite{tanaka2011vocal, midi2008voice}.
Regarding prosody, literature evidences that \ac{PD} patients mainly present a reduction in \ac{SPL} or \ac{SPL} variability \cite{Holmesj2000voice}. They also present a reduced frequency range \cite{Holmesj2000voice,moro2020review} as the most prominent feature.

With the above-mentioned evidence on the characteristics of Parkinsonian speech, literature reports the use of \ac{ML} methodologies to exploit the close relationship between speech and \ac{PD}.
To this respect, automatic speech analysis has shown promising results, providing objective tools to detect and assess the severity of the neurodegenerative effects of the disease \cite{ngo2022computerized} using traditional \ac{ML} techniques \cite{tsanas2012novel} and advanced \ac{DL} approaches. 
In this respect, models using classical voice quality measurements \cite{moro2021advances} or Mel-frequency cepstral coefficients \cite{hawi2022automatic} have shown the potential to model the influence of \ac{PD} on the speech, with recent advances in \ac{DL} offering even greater accuracy by extracting abstract high-level characteristics \cite{vasquez2018multimodal, arias2020predicting, fujita2021performance}. However, training such models requires a broad variety of public datasets of Parkinsonian speech to ensure robustness and generalizability. 

In \ac{PD} speech research, public datasets are notably scarce. Currently, to the author's knowledge, there exists only one public dataset of Italian speakers \cite{dimauro2017assessment}, featuring 831 audio files of 28 patients with \ac{PD} and 50 \ac{HC} speakers. It includes two reading tasks, the repetition of the syllables /pa/ and /ta/, and the sustained phonation of vowels. 
Although this data set provides valuable information, it is limited by its small cohort of \ac{PD} patients, the absence of spontaneous speech, and the lack of manually transcribed speech. 
Furthermore, its reliance on text-dependent utterances read by the participants (i.e., not uttered in an espontaneous way) may pose challenges for elderly participants due to the higher cognitive load required. Regarding Spanish speech and language research, the GITA \cite{orozco2014new} dataset contains data from Colombian Spanish speakers, but its access is restricted and is available only upon request. 
\rw{Regarding the English language, two datasets are well-known. First, the mPower corpus \cite{bot2016mpower}, an American English resource, contains data from self-identified PD and HC individuals. This dataset includes a survey, the MDS-UPDRS questionnaire, various activities, and a voice component featuring 10-second recordings of sustained vowel /a/. With 5,826 participants and 65,022 recordings, mPower is extensive, but its voice data is limited to a single vocal task, which may restrict its use in diverse research contexts. Another relevant dataset is the Mobile Device Voice Recordings recorded by King's College London (MDVR-KCL) \cite{jaeger2019mobile}, which includes recordings from 21 HC individuals and 16 \ac{PD} patients. Each of the 37 participants performed two tasks: reading a paragraph of text and engaging in spontaneous dialogue. The recordings, stored in WAV format, have an average duration of 150 seconds each. However, the small size of this dataset and its limited participant pool highlight the need for more extensive datasets}.

Addressing these significant gaps, \julian{ NeuroVoz (v.3.0.0)} emerges as the most comprehensive publicly available dataset of Parkinsonian speech research to date. In addition, it is the first fully public dataset in Spanish, marking a crucial contribution to the field. NeuroVoz not only increases resources for analyzing PD speech, but also serves as a tool for examining the articulatory, phonatory, and prosodic aspects of Parkinsonian speech. Moreover, this dataset significantly expands the research possibilities within the Spanish-speaking world, embracing the distinct linguistic features of Castilian Spanish and thus enriching the diversity and representativeness of Spanish speech corpora. The NeuroVoz dataset is a specialized collection of speech recordings from individuals diagnosed with \ac{PD} and \ac{HC}, designed to support the development and validation of \ac{ML} models for the diagnosis and monitoring of \ac{PD}.

NeuroVoz was jointly recorded by the \ac{ByO} group from \ac{UPM} and the Otorhinolaryngology and Neurology Services of \ac{HGUGM} and \ac{HUF}, Madrid, Spain. The data set includes a diverse set of speech tasks, ranging from sustained phonation of vowels, text-dependent utterances, a \ac{DDK} test, and a spontaneous free monologue. The specific material recorded is detailed in the Protocol Section. 

The NeuroVoz data set has contributed significantly to research on the screening and detection of \ac{PD} through speech analysis, marked by a series of studies that evolved in both methodology and focus. Moro et al. (2017) \cite{moro2017use} achieved 82\% accuracy using Acoustic Landmarks and a GMM-UBM-Blend classification strategy, emphasizing the utility of burst segments for the screening of \ac{PD}. In a later work, Moro et al. (2018) \cite{moro2018study} introduced a novel approach that combined speaker recognition and allophonic grouping, achieving an accuracy of 86\%. Further studies, such as Moro et al. (2019) \cite{moro2019analysis}, focused on phonatory features that achieved up to 89\% accuracy, while others \cite{moro2019forced, moro2019phonetic} placed the focus on an articulation analysis, reaching accuracies between 81\%-89\%. 
Arias et al. (2020) \cite{arias2020predicting} explored the use of \ac{DL} and transfer learning for \ac{UPDRS} score prediction, highlighting the integration of feature engineering and learning approaches. Furthermore, Godino et al. (2020) \cite{godino2020approaches} underscored artificial intelligence's role in the screening of \ac{PD} using various methods, reporting accuracies over 85\%. 
Lastly, Ibarra et al. (2023) \cite{ibarra2023towards} improved the model generalisability using domain adversarial training, achieving 83\% accuracy in a cross-corpora scenario and Guerrero et al. (2024) \cite{guerrero2024marta} achieved a benchmarking 91$\pm$9 discrimination power. Collectively, these studies establish the NeuroVoz dataset as a valuable benchmark for \ac{PD} detection models, with the current leading performance achieving 89\% accuracy within this dataset, and 83\% for cross-corpora evaluations.

This database, NeuroVoz, marks a significant contribution to the scientific community by offering a comprehensive resource for the systematic exploration of Parkinsonian speech characteristics of Castillian Spanish speakers. It fills a notable gap in the available resources, as currently the only fully public and freely accessible dataset for the analysis of PD speech is in Italian\cite{dimauro2017assessment}, leaving a void for research in Spanish, and more specifically, in Castilian Spanish. By making NeuroVoz publicly available in adherence to the principles of FAIR (Findable, Accessible, Interoperable, and Reusable), we aim to broaden the scope of understanding \ac{PD}'s impact on speech patterns.

\section*{Methods}

A description of several aspects of the corpus is presented next, including the ethics declaration, a description of the participants, the speech protocol tasks, and the audio registration process.

\julian{

\subsection*{Ethics declaration} 

The study was approved by the Ethics Review Board of the \ac{HUF} and \ac{HGUGM} with codes 18/11-ENM1 and 11/2015 respectively, and in accordance with the Spanish Ethical Review Act. All patients and controls were native Spanish speakers and followed the same experimental protocol.

All participants were informed about the project objectives and, if they agreed with the study conditions, they were recruited for participation and recording at \ac{HGUGM} facilities. Participants received a document containing details about the project goals prior to recording. Subsequently, they were asked to sign a consent form. Participants did not receive any compensation for participating in the study, agreeing to share their voices for research purposes. Patients were informed of their rights and the possibility of leaving the study at any time.

Patients were individually identified with a code, which is different from the one used in the Hospital for their clinical histories. No personal data was exchanged with external researchers who had access to the corpus. Only one specialist got in contact with each patient, being also in charge of collecting the clinical data. 

}

\subsection*{Participants}

The dataset comprises 112 individuals, evenly distributed in terms of age and sex, with 46 females, 61 males, and one participant whose sex information was not provided. All participants are native Castillian Spanish speakers.
 
NeuroVoz includes two cohorts: a \ac{PD} group, consisting of patients clinically diagnosed with \ac{PD}; and a \ac{HC} group, comprising normophonic individuals. In total, 58 \ac{HC} speakers and 54 patients with \ac{PD} were recorded.
The mean age of the participants was $71.13 \pm 10.62$ for the \ac{PD} group and $64.04 \pm 10.26$ for the \ac{HC} group. The total number of subjects is limited by the availability of patients received in the Neurology Service of \ac{HGUGM} and \ac{HUF}.

\begin{figure}[ht]
    \centering
    \includegraphics[width=\textwidth]{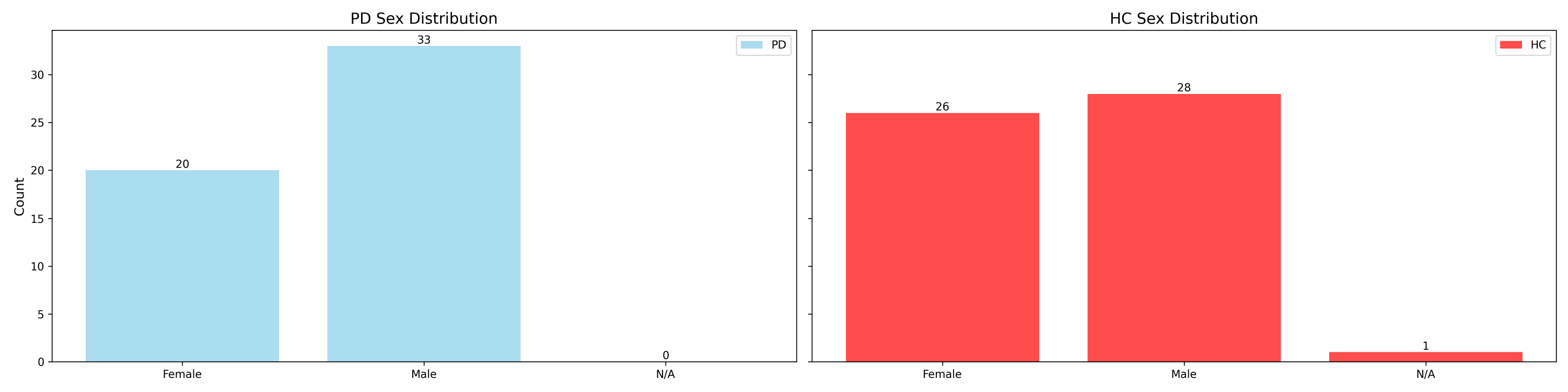}
    \caption{Sex distribution of participants. }
    \label{fig:sexdistro}
\end{figure}

Fig. \ref{fig:sexdistro} shows the sex distribution for the \ac{PD} and \ac{HC} groups, while Fig. \ref{fig:agedistro} presents the age distribution for both cohorts. 
The exclusion criteria for \ac{PD} patients and \ac{HC} participants is included in Table \ref{tab:exclusion_criteria}.

\begin{figure}[ht]
    \centering
    \includegraphics[width=\textwidth]{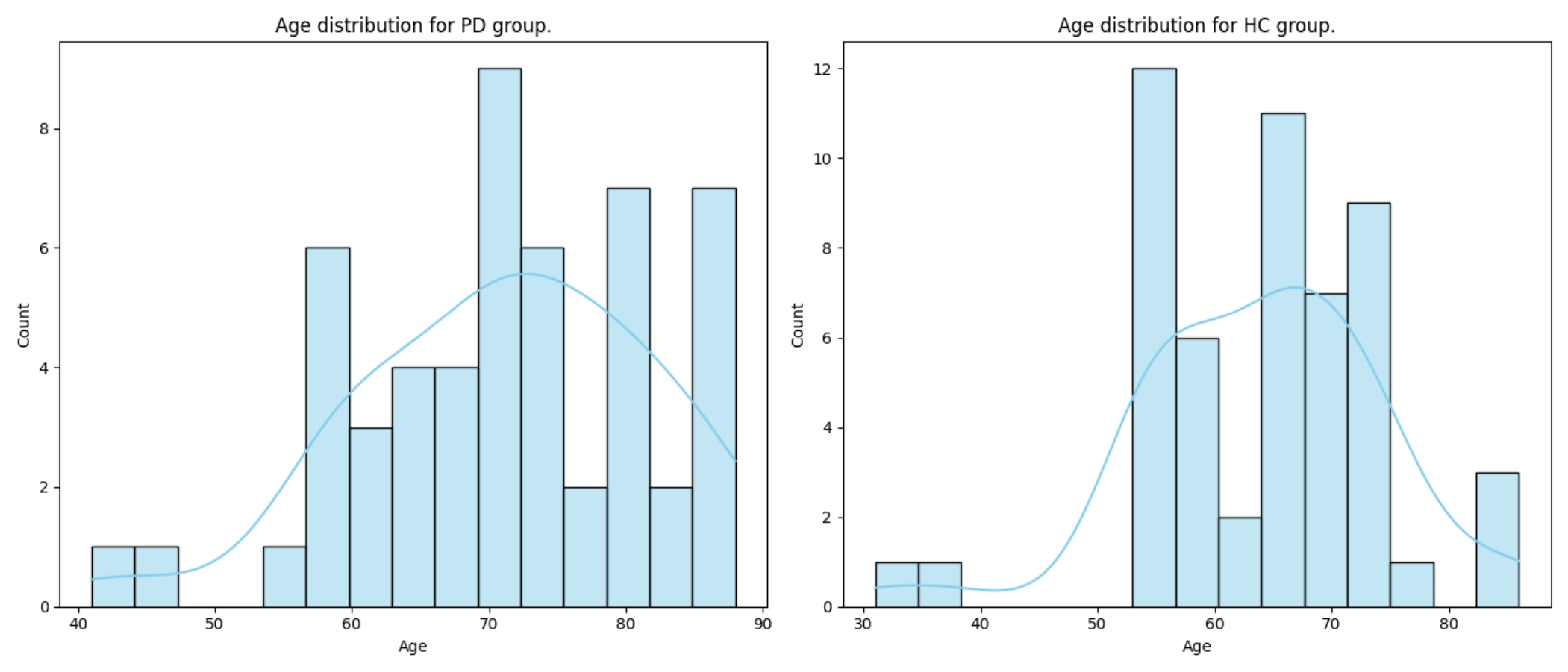}
    \caption{Age distribution of participants.}
    \label{fig:agedistro}
\end{figure}

All participants were evaluated according to the \ac{GRBAS} perceptual scale\cite{hirano1986clinical}, and \ac{PD} patients were also evaluated according to the \ac{UPDRS}\cite{movement2003unified} and \ac{H-Y}\cite{hoehn1967parkinsonism} clinical scales. 

\begin{table}[h]
\centering
\caption{Exclusion Criteria for PD and HC participants.}
\label{tab:exclusion_criteria}
\begin{tabular}{p{0.45\textwidth}|p{0.45\textwidth}}
\hline
\multicolumn{2}{c}{\textbf{Common Exclusion Criteria}} \\ \hline
\multicolumn{2}{c}{A history of voice or speech alterations or laryngeal or vocal cords surgery.} \\
\multicolumn{2}{c}{Regular consumer of alcohol.} \\
\multicolumn{2}{c}{Smoker.} \\
\multicolumn{2}{c}{Speaker whose mother tongue is not Spanish.} \\
\multicolumn{2}{c}{Speaker suffering from any disease that affects speech intelligibility or affects speech or voice.} \\ \hline
\textbf{PD Patients} & \textbf{HC Participants} \\ \hline
Diagnosis of another neurological disease. & Participants with neurological disorders and/or PD. \\ \hline
\end{tabular}
\end{table}

To ensure the exclusion of organic vocal pathologies (i.e. lesions, scar, reflux) and facilitate a comprehensive clinical diagnosis, an otolaryngologist also performed an evaluation of symptoms such as vocal tremor, cephalic tremor, mandibular tremor, siallorrhea, dysphagia, and hipophonic voice.

The inclusion criteria for Parkinsonian speakers included patients diagnosed with \ac{PD} by the \ac{HGUGM} and \ac{HUF} Neurology Services, no matter the start date of clinical signs. Participants with \ac{PD} were recorded in ON state, having taken the prescribed medication -when applicable- 2 to 5 hours before the recording session. Most patients have been diagnosed with \ac{PD} for less than 10 years, as seen in Fig. \ref{fig:PD_distribution}.

\begin{figure}[ht]
    \centering
    \includegraphics[width=0.5\textwidth]{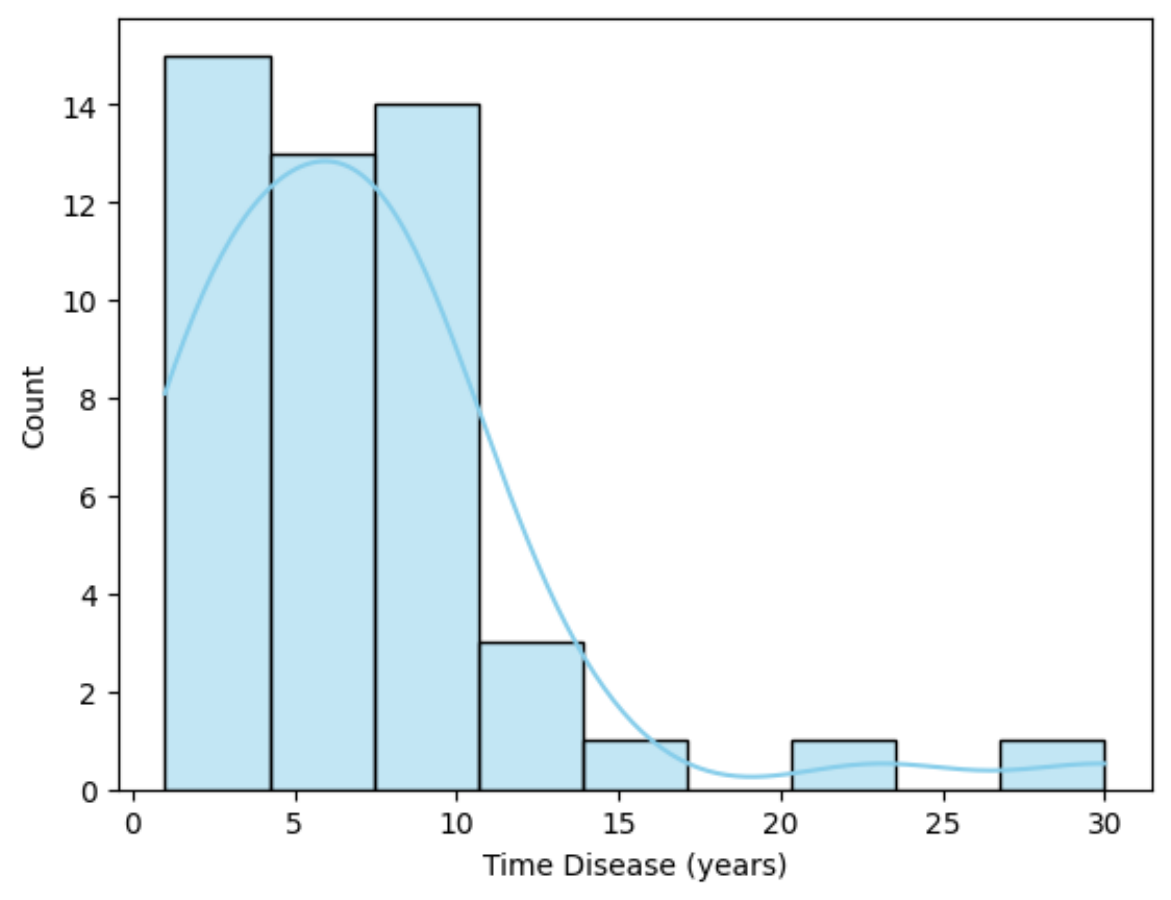}
    \caption{Distribution of years since the diagnosis of \ac{PD}.}
    \label{fig:PD_distribution}
\end{figure}

For the \ac{HC} group, volunteers were recruited based on the following criteria: they were not professional voice users (i.e., their voice is not essential for their job); had no neurological diseases; and they had no diagnosis of \ac{PD}. In addition, the \ac{HC} group was selected to match Parkinsonian patients in terms of sex and age. 
The \ac{HC} group was also subjected to a neurological state assessment by a survey.

The distribution of \ac{GRBAS} total scores, i.e. sum up all GRBAS values for the same patient for each audio, are shown in Fig. \ref{fig:grbasdistro}.

\begin{figure}[ht]
    \centering
    \includegraphics[width=\textwidth]{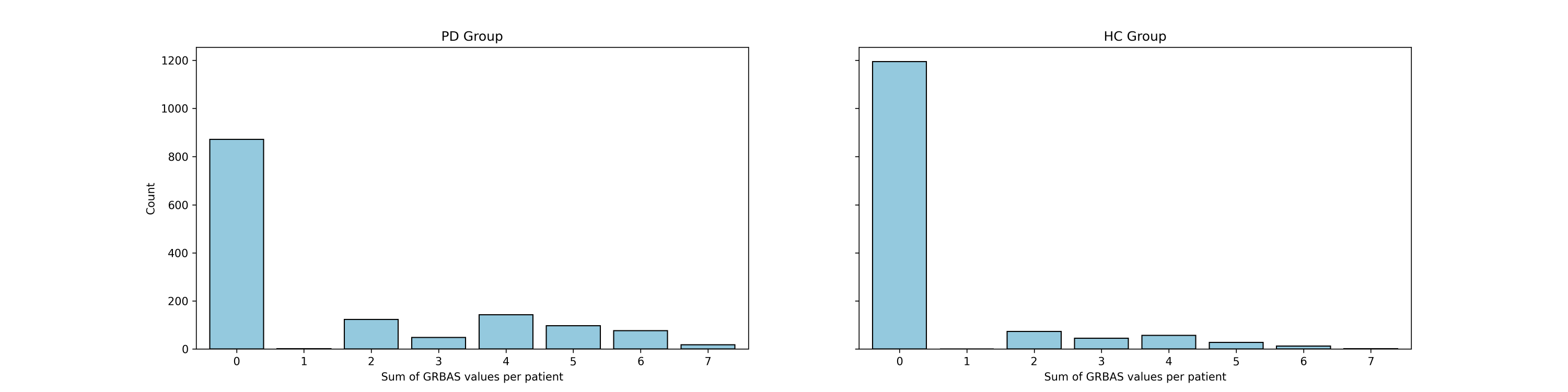}
    \caption{GRBAS total sum distribution of participants.}
    \label{fig:grbasdistro}
\end{figure}

\subsection*{Recording protocol}\label{cap:Protocol}

\rw{PD is a neurodegenerative disease causing movement disorders and, thus, the effect of PD is initially expected to be mother tongue independent. But articulations during speech production are slightly different for each language (or even dialect), due to differences mainly attributed to phonetic, phonological and linguistic variances. 
Thus, the effect of PD in the speech is expected to introduce small differences that are language and/or dialect dependent. Besides, certain manner classes are considered more relevant than others for a good characterization of the effect of the disease \cite{moro2019phonetic}. Thus, the recording protocol should pay attention to be phonetically balanced, but also to include a wide variety of the most affected phonemes by the disease. To this respect, literature reports that plosives are very relevant for the detection of PD, but also fricatives \cite{moro2019phonetic}.} 

In view of the afforementioned, an extensive recording protocol was designed. The protocol was developed to facilitate a further detailed analysis of known deviations introduced by the disease with respect to the expected normality. 

The protocol includes four different speech tasks:

\begin{itemize}
    \item \textbf{Sustained phonation of vowels:} Participants were instructed to phonate the five Spanish vowels /a/, /e/, /i/, /o/, and /u/ at a comfortable pitch and volume, taking brief pauses to breathe between each phonation. This procedure was repeated three times. The mean duration of sustained vowel phonation was $3.89 \pm 0.73$ s. for the \ac{PD} group and $4.38 \pm 1.54$ s. for the \ac{HC} group.
    
    \item \textbf{\ac{LR} sentences:} Participants were instructed to listen to 16 predefined phrases and repeat them aloud. This method, chosen over reading a text, aimed to lessen the cognitive load of the process for the patients. Sentences are familiar sayings that are easy to recall due to their recognizability. Additionally, this method mitigates potential biases in speech that could arise during reading due to vision impairments common among the elderly population, thereby facilitating a more natural and impartial articulation process. The predefined phrases are detailed in Table \ref{tab:IPATDU}. The mean duration of the sentences was $11.11 \pm 4.64$ s. for the \ac{PD} group, and $13.06 \pm 5.46$ s. for the \ac{HC} group. These sentences were  chosen to be acoustically balanced in conjunction, but also to ensure a wide variety of plosive and fricative sounds, and to force frequent velum openings and closures. 
    
    \item \textbf{\ac{DDK} test:} Participants were instructed to perform the repetition of the sequence /pa-ta-ka/ during, at least, 5 s., and as fast as possible. The mean duration of the DDK tests were $3.26 \pm 2.45$ s. for the \ac{PD} group and $3.70 \pm 2.74$ s. for the \ac{HC} group. 
    
    \item \textbf{Monologues:} Participants were also asked to deliver a free monologue, describing a predetermined illustration that depicts various activities of daily living, as illustrated in Fig. \ref{fig:monologue}. This task was designed to gather spontaneous speech, further enriching the dataset with more natural speech patterns. The average duration of these monologues was $31.15 \pm 12.43$ s. for the \ac{PD} group, and $46.53 \pm 21.41$ s. for the \ac{HC} group.
\end{itemize}

\begin{figure}
\centering
\includegraphics[width=0.3\textwidth]{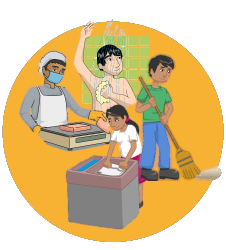}
\caption{The free monologue is guided by asking the patient to describe the scenes presented in this illustration.}
\label{fig:monologue}
\end{figure}

The order followed to perform the speech recordings is detailed in Table \ref{tab:speech_task_protocol}. Initially, for a further evaluation of the phonation, each sustained vowel was recorded independently. In the second place, eight different \ac{LR} phrases were recorded to assess the velopharyngeal closure, which is often associated with Parkinsonian speech \cite{hoodin1989parkinsonian}. Subsequently, the patient underwent the DDK test, which is typically used to evaluate the articulation ability. Next, three additional \ac{LR} phrases were recorded for a potential analysis of prosodic aspects. Intonation and emotion were addressed by asking the patient to emphasize certain words in a more emotional manner (represented with capital letters in Table \ref{tab:IPATDU}). Finally, three more \ac{LR} phrases were recorded.

\begin{table}[h]
\centering
\caption{Speech tasks protocol followed in order}
\label{tab:speech_task_protocol}
\begin{tabular}{p{0.1\textwidth}p{0.2\textwidth}p{0.6\textwidth}}
\hline
\textbf{Order} & \textbf{Task} & \textbf{Description} \\ \hline
1 & Phonation & Repeating sustained vowels to evaluate phonation. \\ 
2 & Velopharyngeal Closure & \ac{LR} phrases 1, 2, 6, 9, 10, 11, 12 and 13 from Table \ref{tab:IPATDU} for velopharyngeal closure assessment. \\ 
3 & Articulation & DDK test to evaluate articulation capabilities. \\ 
4 & Prosody & \ac{LR} phrases 3 and 7 from Table \ref{tab:IPATDU} for prosody evaluation. \\ 
5 & Intonation-Emotion & \ac{LR} phrases 5, 14, and 15 from Table \ref{tab:IPATDU} with emphasis on capitalised words for intonation-emotion assessment. \\ 
6 & Other tasks & \ac{LR} phrases 4, 8, and 10 from Table \ref{tab:IPATDU}. \\ \hline
\end{tabular}
\end{table}

Table \ref{tab:IPATDU} specifies the phrases used for the \ac{LR} tasks along with their translation to English and their \ac{IPA} transcriptions. 

\begin{table*}[ht]
\centering
\caption{Transcriptions in the \ac{IPA} and translations of selected sentences}
\label{tab:IPATDU}
\begin{tabular}{ p{0.5cm} p{2.3cm} p{4.3cm} p{4.3cm} p{4.3cm} }
\toprule
\textbf{Sent. \#} & \textbf{ID} & \textbf{Spanish transcription} & \textbf{IPA transcription} & \textbf{English translation} \\ \midrule
1 & ABLANDADA &  \textit{La patata no está bien ablandada} & [\textipa{la pa"tata no "esta Bjen aBlan"dada}] & ``The potato is not soft enough'' \\
2 & ACAMPADA & \textit{Mañana vamos de acampada} & [\textipa{ma"26ana "Bamos de akam"pada}] & "Tomorrow we are going camping" \\
3 & BARBAS & \textit{Cuando las barbas de tu vecino veas pelar, pon las tuyas a remojar} & [kwando las \textipa{BarBas} \textipa{de} tu \textipa{Be\texttheta ino} \textipa{Beas} pelar pon las \textipa{tu\textipa{J}as} a \textipa{remo\textipa{X}ar}] & "When your neighbor's beard you see peeling, put yours to soak" \\
4 & BURRO & \textit{Burro grande ande o no ande} & [\textipa{"bur.o "gran.de "an.de o no "an.de}] & "Big donkey, walk or not walk" \\
5 & CALLE & \textit{De la calle VENDRÁ quien de tu casa te ECHARÁ} & [de la \textipa{ka\textipa{Je}} \textipa{Bend\textipa{Ra}} \textipa{kj\textipa{en}} \textipa{de} tu \textipa{kasa} te e\textipa{tSa}\textipa{ra}] & "From outside will come who will kick you out from your house" \\
6 & CARMEN & \textit{Carmen baila el mambo} & [\textipa{"kar.men "bai.la el "mam.bo}] & "Carmen dances the mambo" \\
7 & DIABLO & \textit{Cuando el diablo no sabe qué hacer, con el rabo mata moscas} & [kwando el \textipa{\textdyoghlig}\textipa{a\textbeta lo} no \textipa{sa\textipa{Be}} ke a\textipa{\texttheta}er kon el \textipa{ra\textbeta o} mata \textipa{moskas}] & "When the devil does not know what to do, it kills flies with its tail" \\
8 & GANGA & \textit{Esto es una ganga} & [\textipa{"es.to es "u.na "gaN.ga} & "This is a bargain" \\
9 & MANGA & \textit{Juan tira de la manga} & [\textipa{"xwan "ti.ra de la "maN.ga}]  & "Juan pulls the sleeve" \\
10 & PERRO & \textit{Dame pan y llámame perro} & [\textipa{"da.me pan i ja.ma.me "pero}] & "Give me bread and call me dog" \\
11 & PAN & \textit{Al pan pan y al vino vino} & \textipa{[al pan pan i al \textbeta ino \textbeta ino]} & "To the bread, bread, and to the wine, wine
" \\
12 & PATATA & \textit{La patata blanda es buena} & \textipa{[la pa"tata "blanda es "bwena]} & "The soft potato is good" \\
13 & PETACA & \textit{La petaca blanca es mía} & [\textipa{[la pe"taka "blanka es "mi.a]}] & "The white flask is mine" \\
14 & PIDIO & \textit{No pidas a quien PIDIÓ, ni sirvas a quien SIRVIÓ} & [no \textipa{pi\textipa{Das}} a \textipa{kj\textipa{en}} \textipa{pi\textipa{\textdyoghlig}\textipa{o}} ni \textipa{si\textipa{R}\textipa{Bas}} a \textipa{kj\textipa{en}} \textipa{si\textipa{R}\textipa{B}\textipa{Jo}}] & "Do not beg the one who begged, nor serve the person who served" \\
15 & SOMBRA & \textit{El que a BUEN árbol se arrima, BUENA sombra le cobija} & [\textipa{el ke a \textbeta wen "a\textfishhookr \textbeta ol se a"rima, "bwena "somb\textfishhookr a le ko"\textbeta ixa}
] & "Who leans close to a good tree is sheltered by good shade" \\
16 & TOMAS & \textit{Tomás tira de la manta} &[\textipa{to"mas "ti.ra de la "man.ta}] & "Tomás pulls the blanket" \\ \bottomrule
\end{tabular}
\end{table*}

The audio was manually post-processed and edited to remove background noise, coughing, and long silences. Moreover, if the audio quality was poor, it was removed from the dataset, Fig. \ref{fig:missing_audios} shows the completeness of each audio material for all patients.

\begin{figure}[ht]
    \centering
    \includegraphics[width=\textwidth]{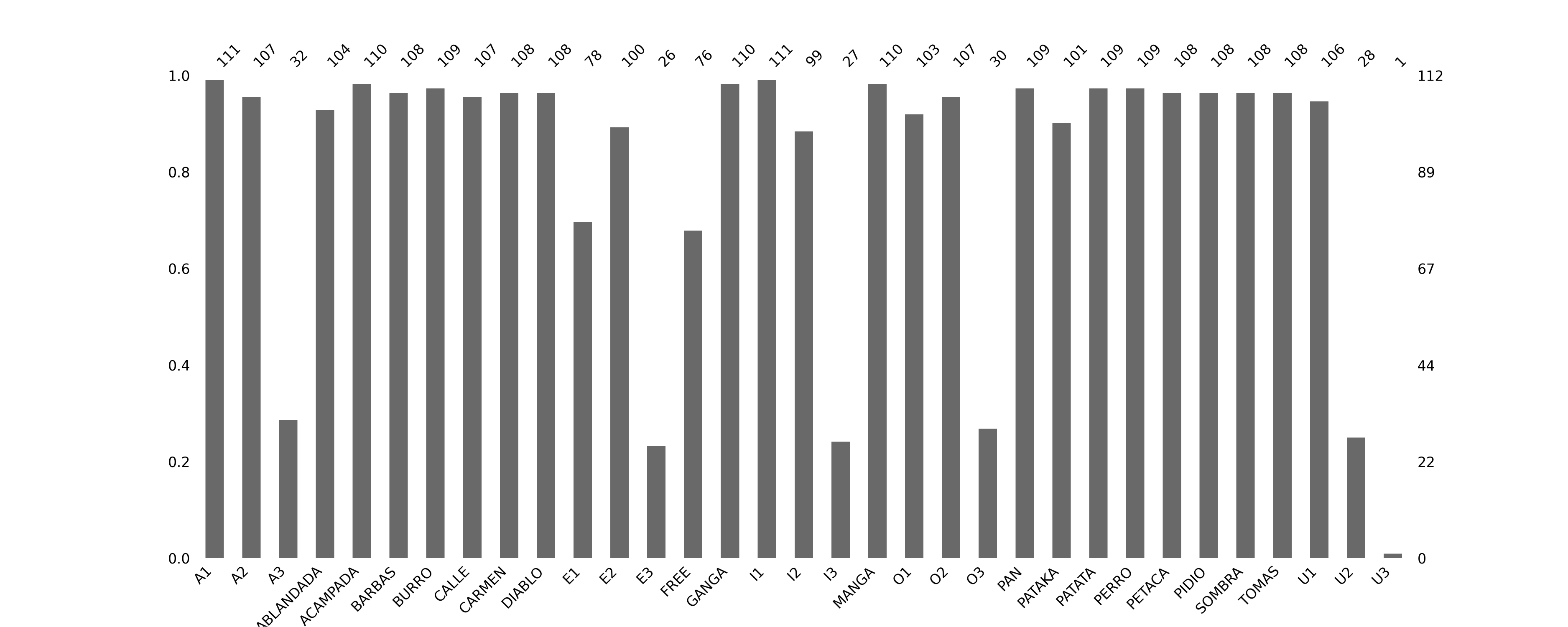}
    \caption{Percentage of completeness of each audio material for all patients.}
    \label{fig:missing_audios}
\end{figure}

\subsection*{Recording setup}

The audio recording setup was standardised to ensure high-quality data suitable for further analysis. 

All participants were recorded under the same environmental conditions and using the same equipment. The technical setup for the audio recordings consisted of an AKG\textsuperscript{®} C420 headset microphone. This microphone was connected to a preamplifier equipped with phantom power, ensuring optimal signal quality. The preamplifier was attached to a SoundBlaster\textsuperscript{®} Live sound card, which operates at a sampling rate of 44,100 Hz. The recording system achieved a mean \ac{SNR} of 24.3 dB, indicating a high-quality audio capture. The audio device was connected to a personal computer running MedivozCaptura\textsuperscript{®} software \cite{godino2006integrated}. The recording was carried out in a quiet room during the clinical routine, but in a non-acoustically isolated environment. 

Throughout the recording sessions, the patient's comfort was prioritized. The participants were instructed to speak in their natural pitch and with a comfortable loudness. Pristine attention to the recording conditions not only ensures the comfort of the participants, but also contributes to the consistency and reliability of the collected data.

The audio files were also carefully post-processed and edited to exclude interjections, coughing, and background noise during silence periods.

\subsection*{Metadata associated to the speaker}

In addition to speech recordings, demographic and clinical data were systematically collected for each participant at the beginning of each session. These data are summarized in Table \ref{tab:metadata}. The date of the evaluation and recording was rounded to the first day of the corresponding month to avoid a potential identification of the speaker by crossing information.

\begin{table}[h]
\centering
\caption{Summary of the metadata collected for each participant.}
\label{tab:metadata}
\begin{tabular}{ll}
\hline
\textbf{Data Type} & \textbf{Description} \\ \hline
Date of Recording & Date \\
Sex & Free text string \\
Age & Integer value \\
Diagnosis & Free text field \\
Symptoms & Binary variables (True/False): \\
& \quad - Vocal tremor \\
& \quad - Cephalic tremor \\
& \quad - Mandibular tremor \\
& \quad - Sialorrhea \\
& \quad - Dysphagia \\
& \quad - Hypophonic voice \\
Duration of disease & Integer value (years) \\
URPDS & \ac{UPDRS} scale \\
\ac{H-Y}  & \ac{H-Y} scales \\
Date of evaluations  & Date \\
Medication & Free text field \\
Medical treatment during recording & Annotated "ON"/"OFF" \\
Occupation & Free text string \\
Vocal folds analysis & String field (Normal/Not performed/free text) \\
Observations & Free text field for additional notes \\ \hline
Doctor & Categorical value to indicate who evaluated the speaker \\
\end{tabular}
\end{table}

The sex is stored as a float where "1" indicates male, "0" female and "NaN" for non-declared cases. Age was recorded as an integer. Moreover, the participants' employment status was also documented, revealing that 63.88\% of the total 112 individuals were retired, which is in consonance with the demographic composition of the dataset.

The diagnostic data identified by the physician included two different diagnoses, being Idiopathic \ac{PD} the most prevalent (affecting 50 patients) and Parkinsonian Syndrome the less frequent (affecting the remaining 5 patients). An extra field contains medical annotations about the diagnosis, such as the most affected body part.

Binary indicators were used to annotate symptoms such as vocal tremor, cephalic tremor, mandibular tremor, sialorrhea, dysphagia, and hypophonic voice, with "1" representing presence and "0" absence, as shown in Fig. \ref{fig:binarydata}. In case of missing data, a float "NaN" is found. Medication details, when applicable, were recorded in a free text field. The time since diagnosis (in years) was also recorded. 

\begin{figure}[ht]
 \centering \includegraphics[width=\textwidth]{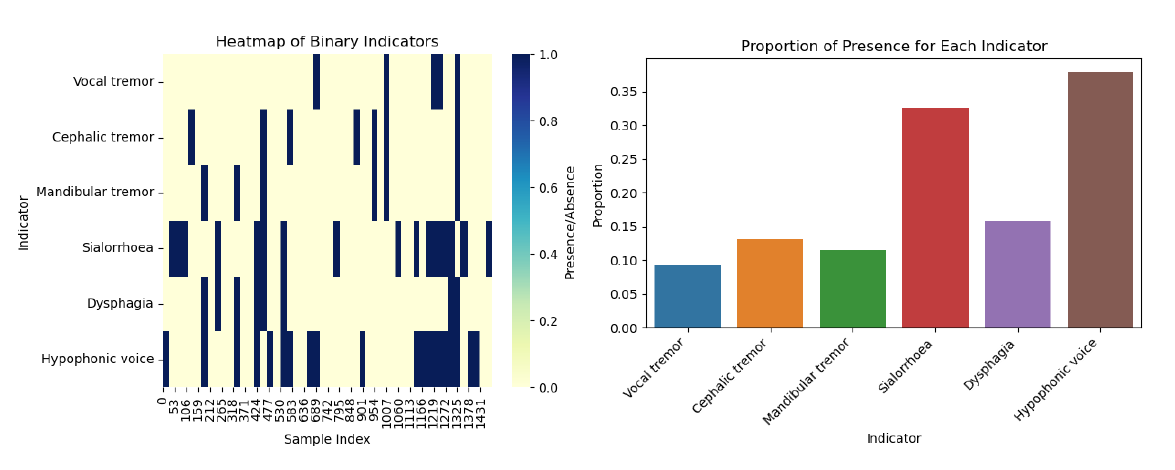}
    \caption{Binary indicators regarding parkinsonian symptoms. Left: Binary indicators of the presence of symptoms for each \ac{PD} patient. The horizontal axis determines the patient ID while the vertical axis denotes each symptom. Right: Binary proportion of presence of symptoms for each \ac{PD} patient.}
    \label{fig:binarydata}
\end{figure}

Clinical evaluations were carried out following the \ac{UPDRS} scale \cite{movement2003unified}, and \ac{H-Y} staging system \cite{hoehn1967parkinsonism}. \ac{UPDRS} was developed by the \ac{MDS} to evaluate various aspects of \ac{PD} including non-motor and motor experiences of daily living and motor complications. It characterizes the extent and burden of disease ranging from 0 and 260. On the other hand, the \ac{H-Y} scale is used to characterize the progression of the disease, classifying the patient according to the severity of the symptoms ranging from 0 to 5. The date on which the patients were evaluated was annotated and annonymized to the first day of the month.
Both clinical evaluations were carried out by four different physicians, all of them certified by the \ac{MDS} for this purpose. Fig. \ref{fig:severitydistro} illustrates a histogram of the results of these evaluations. 

\begin{figure}[ht]
    \centering
    \includegraphics[width=\textwidth]{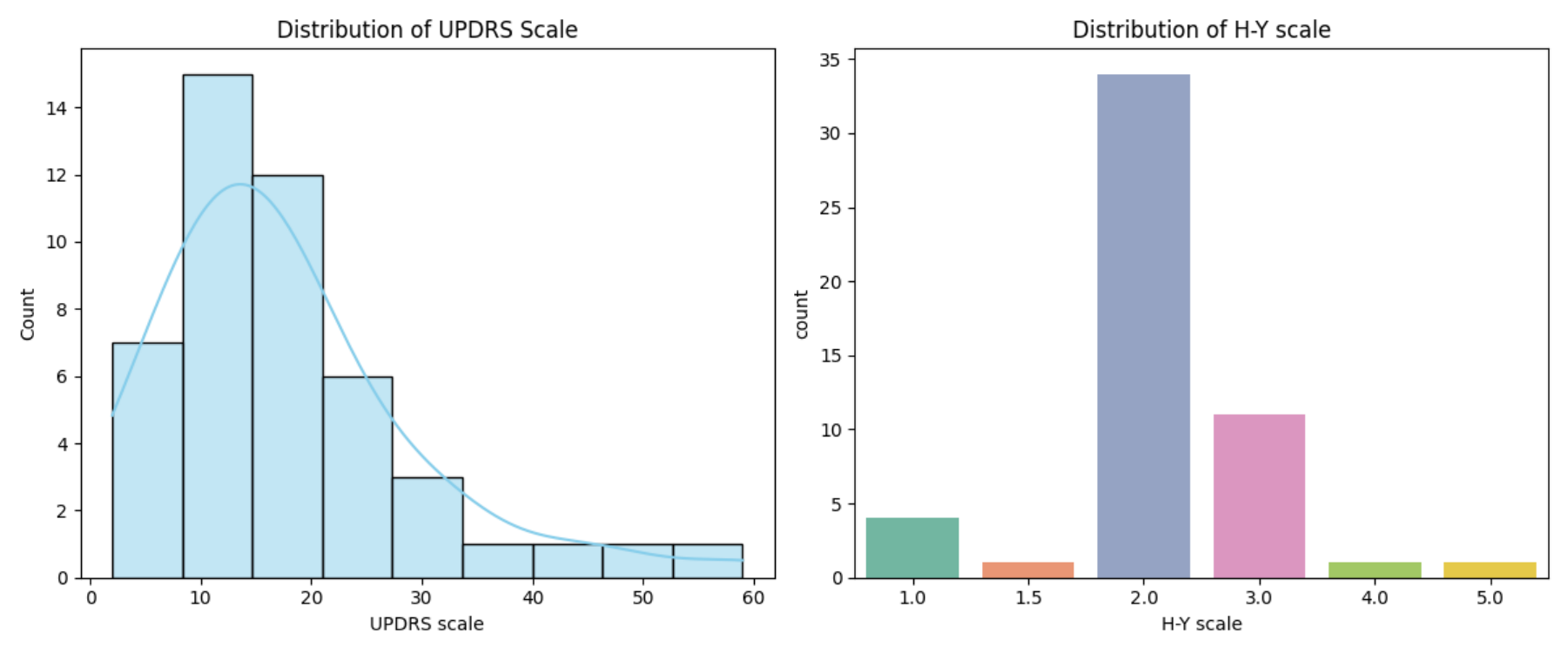}
    \caption{Severity distribution of parkinsonian patiens expressed in \ac{UPDRS} and \ac{H-Y} scales.}
    \label{fig:severitydistro}
\end{figure}

The findings of the direct examination of the vocal folds using fiber-laryngoscopic methods were also documented, revealing that all participants except one \ac{HC} had normal vocal folds (this participant had a hypophonic right vocal corone). In any case, it is worth noting that a relevant proportion of participants declined this examination —28.30\% of the \ac{PD} group, and 38.11\% of the \ac{HC} group.

\subsection*{Metadata associated to the audio recordings}

Each recording also has complementary information associated, such as the audio manual transcription, the \ac{GRBAS} perceptual evaluations, or a set of voice quality characteristics. The metadata associated are dependent on the specific acoustic material in the recording, being different for vowels, monologues, or \ac{LR} eloquences. 

\subsubsection*{Audio transcriptions}

To ensure accurate speech annotation, manual transcriptions were meticulously performed for both \ac{LR} eloquences and monologues. This is a critical task, especially for the \ac{LR} eloquences. The goal was to identify articulation mistakes made by the speakers or linguistic discrepancies between the utterances produced by the speakers and those expected. Trained transcribers carefully listened to each audio, recording the transcriptions word by word, and noting any deviations or mispronunciations. 

\subsubsection*{Voice quality features}

A set of voice quality features, detailed in Table \ref{tab:audio-parameters}, was calculated and included in the corpus. These features were extracted for the three repetitions of all sustained vowels. The AVCA-ByO open-source package \cite{gomez2021design} was used to extract three subsets of characteristics, namely: i) amplitude and frequency perturbation features; ii) noise parameters; and, iii) tremor parameters.

\begin{table}[ht]
\centering
\caption{Voice measurements extracted from the recordings of the /a/ sustained vowels.}
\label{tab:audio-parameters}
\begin{tabular}{lcc}
\toprule
\textbf{Parameter} & \textbf{Abbreviation} & \textbf{Unit Measure} \\
\midrule
\multicolumn{3}{c}{\textbf{Perturbation Measures}} \\
\midrule
Absolute Jitter & Jitter & $\mu$Seconds \\
Relative Jitter & rJitter & \% \\
 Relative Average Perturbation & RAP & \% \\
Pitch Period Perturbation Quotient & rPPQ & \% \\
 Smoothed Pitch Period Perturbation Quotient & rSPPQ & \% \\
Absolute Shimmer & ShimmerDb & dB \\
Relative Shimmer & rShimmer & \% \\
 Amplitude Perturbation Quotient & APQ & \% \\
Smoothed Amplitude Perturbation Quotient & sAPQ & \% \\
Cepstral Peak Prominence & CPP & dB \\
\midrule
\multicolumn{3}{c}{\textbf{Noise Parameters}} \\
\midrule
 Harmonics-to-Noise Ratio & HNR & dB \\
 Cepstrum Harmonics-to-Noise Ratio & CHNR & dB \\
 Glottal to Noise Excitation Ratio & GNE & Ratio \\
 Normalised Noise Energy & NNE & dB \\
\midrule
\multicolumn{3}{c}{\textbf{Tremor Parameters}} \\
\midrule
Frequency Tremor Intensity Index & FTRI & Arbitrary Units \\
Amplitude Tremor Intensity Index & ATRI & Arbitrary Units \\
Fundamental Frequency Tremor Frequency & FFTR & Hz \\
Amplitude Tremor Frequency & ATRF & Hz \\
\bottomrule
\end{tabular}   
\end{table}

\subsubsection*{Perceptual evaluations}

The \ac{GRBAS} perceptual scale \cite{hirano1986clinical}, is a globally recognized tool for vocal assessment \cite{bhuta2004perceptual, karnell2007reliability, kojima2021objective}, providing a structured method for clinicians and researchers to evaluate dysphonia \cite{isshiki1969differential}. It is both reliable and practical, posing no discomfort or inconvenience to either the patient or the evaluator \cite{nemr2012grbas}. Each \ac{GRBAS} element is scored with a unique value, and a cumulative score combining all is also obtained. The distribution of these evaluations for each group are detailed in Fig. \ref{fig:grbas_dsitro}.

\begin{figure}
    \centering
    \includegraphics[width=\textwidth]{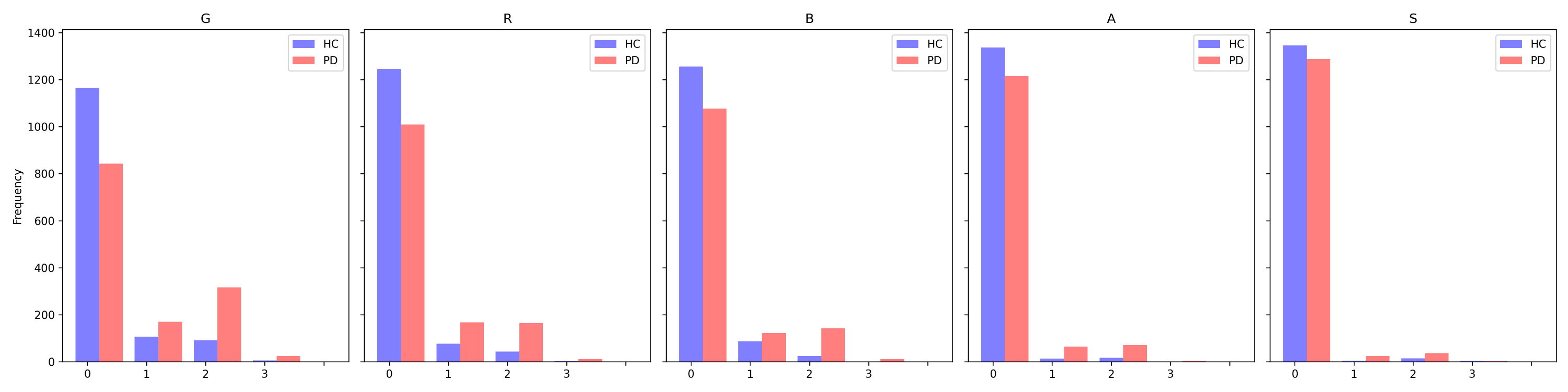}
    \caption{Distribution of GRBAS individual evaluations for HC and PD patients}
    \label{fig:grbas_dsitro}
\end{figure}

A speech therapist with more than 15 years of professional experience evaluated each audio using the scale, scoring the five different elements of vocal quality. Grade (G) (the overall severity of dysphonia), Roughness (R), Breathiness (B), Asthenia (A) (vocal weakness), and Strain (S). The evaluation consists in assigning an integer value ranging from 0 to 3 to each element of the scale, where "0" represents normal, and "1", "2", and "3" indicate slight, moderate, and severe abnormalities, respectively. 

For the sustained vowels, a \ac{GRBAS} rating was assigned to every recording, along with additional comments from the speech therapist. For \ac{LR} tasks and spontaneous monologues, the \ac{GRBAS} evaluation was supplemented with an additional perceptual voice analysis, including an evaluation of the glottal attack (rated as normal, soft, hard, or somewhat hard), tone (normal, high-pitched, deep, or unstable), and phonation quality (normal, forced-strangulated, breathy, wet, or blown). Intensity levels were categorized as normal, weak, or increased, while speech rate was evaluated as normal, slow, or accelerated. Resonance was described as normal, hypernasal, or hyponasal. Intelligibility was graded on a scale from 1 (normal) to 5 (severe deficiency), and speech prosody was assessed as normal, monotonous, variable intensity, or excessive accentuation. Lastly, articulation was evaluated as normal, slurred, or exaggerated.

\section*{Data Records}

Data access is facilitated through the Zenodo repository \cite{neurovoz_zenodo}. The structural representation of the data set can be seen in Fig. \ref{fig:db-tree-st}. The total size of the data set is 1.01 GB, comprising a collection of 2,977 audio files. On average, each patient contributes approximately with $26.88 \pm 3.35$ audio recordings.

FIGURE 10 GOES HERE

The data is distributed into five distinct folders, namely, \texttt{"audio features"}, \texttt{"audios"}, \texttt{"grbas"}, \texttt{"transcriptions"} and \texttt{"metadata"}. In the subsequent sections, a detailed overview of the contents of each of these folders is provided.

\subsection*{Voice quality features}\label{cap:AudioFeatures}
The \texttt{"audio features"} folder contains all precomputed voice quality features. These features are available exclusively for sustained vowels. These voice quality features are stored in a \texttt{.csv} file named \texttt{audio\_features.csv}. The \texttt{.csv} file comprises 18 columns: the initial column indicates the path to the audio file, while the remaining 17 columns represent each of the precomputed voice quality features.

\subsection*{Audio recordings}

The \texttt{"audios"} directory serves as the repository for all audio recordings, each stored in the \texttt{.wav} format. The naming convention of the audio files encapsulates critical information about the patient's condition, the specific audio material recorded, and the unique identifier of the patient, structured as follows:

"\texttt{\textbf{Condition}\_\textbf{AudioMaterial}\textbf{\#Repetition}\_\textbf{IDPatient}.wav}".

This format is deciphered as follows:

\begin{itemize}
    \item \texttt{\textbf{Condition}}: Identifies the patient's health status, categorized as either \ac{HC} or \ac{PD}, according to the health condition.
    \item \texttt{\textbf{AudioMaterial}}: Denotes the type of acoustic material, which may be "A", "E", "I", "O", or "U" for the sustained vowels; a keyword (as detailed in Table \ref{tab:IPATDU}) for the text dependent utterances; "FREE" for free speech monologues; or "PATAKA" for the \ac{DDK} test.
    \item \texttt{\textbf{\#Repetition}}: Applicable only to the sustained vowels -performed three times-, indicating the sequence of the attempt (ranging from 1 to 3).
    \item \texttt{\textbf{IDPatient}}: A four-digit integer uniquely designating the patient.
\end{itemize}

For instance, the filename \texttt{\textbf{HC}\_\textbf{A}\textbf{2}\_\textbf{0034}.wav} represents the recording of the second attempt of the sustained vowel /a/ by patient 0034, who belongs to the \ac{HC} cohort. Similarly, \texttt{\textbf{PD}\_\textbf{GANGA}\_\textbf{0066}.wav} corresponds to the recording of the phrase tagged as "GANGA" in Table \ref{tab:IPATDU} for patient 0066, belonging to the  \ac{PD} group.

\subsection*{GRBAS evaluations}
The \texttt{"grbas"} folder contains all \ac{GRBAS} evaluations performed by a speech therapist. These data are organized into 32 distinct \texttt{.csv} files, each named according to the \textbf{\texttt{AudioMaterial}} identifier.

For recordings of sustained vowels, such as \texttt{A1.csv} for the first phonation of /a/ or \texttt{E3.csv} for the third phonation of /e/, the files feature 8 columns. These include: the filename, indicating the \texttt{.wav} file's name; five columns,  holding an integer score with the evaluation corresponding to each of the five \ac{GRBAS} features; a column for the overall \ac{GRBAS} score; and a column to record the observations of the therapist about the recording.

Regarding other audio tasks, such as \ac{LR} or DDK tasks, different \texttt{.csv} are found. These \texttt{.csv} files, for example \texttt{Barbas.csv}, contain 16 columns. The initial 8 mirror those for the vowel assessments, detailing filenames, \ac{GRBAS} partial scores, total \ac{GRBAS} score and comments. The subsequent eight columns provide the therapist's perceptual notes, including glottal attack (classified as normal, soft, hard, or somewhat hard), tone (normal, high-pitched, grave, or unstable), and quality of phonation (normal, forced-strangulated, breathy, wet, or trembling). Intensity is labelled as normal, weak, or increased; speech pace as normal, slow, or fast; resonance as normal, hypernasal, or hyponasal. Intelligibility ranges from normal to severe deficiency. Prosody is assessed for normality, monotony, intensity variations, or excessive emphasis. And articulation is evaluated as normal, slurred, or overstated.

\subsection*{Transcriptions}

The \texttt{"transcriptions"} directory serves as the repository for all audio manual transcriptions, each stored in a \texttt{.txt} file encoded in latin-1. The naming convention of the text files follows the same rules as the audio files, structured as follows: 

"\texttt{\textbf{Condition}\_\textbf{AudioMaterial}\_\textbf{IDPatient}.txt}".

This format is deciphered as:

\begin{itemize}
    \item \texttt{\textbf{Condition}}: Identifies the patient's health status, categorized as either \ac{HC} or \ac{PD}, reflecting the health condition.
    \item \texttt{\textbf{AudioMaterial}}: Denotes a keyword from specific phrases (as detailed in Table \ref{tab:IPATDU}) or "FREE" for free speech monologues.
    \item \texttt{\textbf{IDPatient}}: A four-digit integer uniquely designating the patient.
\end{itemize}

For instance, the filename \texttt{\textbf{HC}\_\textbf{ABLANDADA}\_\textbf{0034}.txt} represents the text transcription of the first sentence in Table \ref{tab:IPATDU} by patient 0034, who falls under the\ac{HC} category. Similarly, \texttt{\textbf{PD}\_\textbf{GANGA}\_\textbf{0066}.txt} corresponds to the transcription of the audio tagged as "GANGA" in Table \ref{tab:IPATDU} for patient 0066, identified with \ac{PD}.

\subsection*{Other metadata}

The \texttt{"metadata"} folder contains the clinical data and metadata recorded in addition to the audio files. It comprises two \texttt{.csv} files, one for each condition (i.e., \ac{HC} and \ac{PD}) named \texttt{metadata\_hc.csv} and \texttt{metadata\_pd.csv}.

Each \texttt{.csv} file is structured around 23 columns: the initial column represents the patient's ID; the second column specifies the cohort ("HC" or "PD"); and the subsequent 10 columns the detailed clinical data as outlined in the methodology. The last column contains the file path to the corresponding audio recording.

\section*{Technical Validation}

The NeuroVoz corpus has been rigorously examined in several studies \cite{moro2017use, moro2018study, moro2019analysis, moro2019forced, moro2019phonetic, arias2020predicting, godino2020approaches, ibarra2023towards}, establishing its usefulness to differentiate between parkinsonian and normophonic speech with a benchmark accuracy of $89 \pm 7$ reported in \cite{moro2019phonetic} using \ac{LR} tasks.

Data integrity and completeness are visually represented in Fig. \ref{fig:missigness}, providing a clear view of the missing data, and reinforcing the dataset's structured approach towards capturing distinct participant groups, while notably, \ac{PD}-specific attributes are absent in the \ac{HC} group.

\begin{figure}
 \centering
    \includegraphics[width=\textwidth]{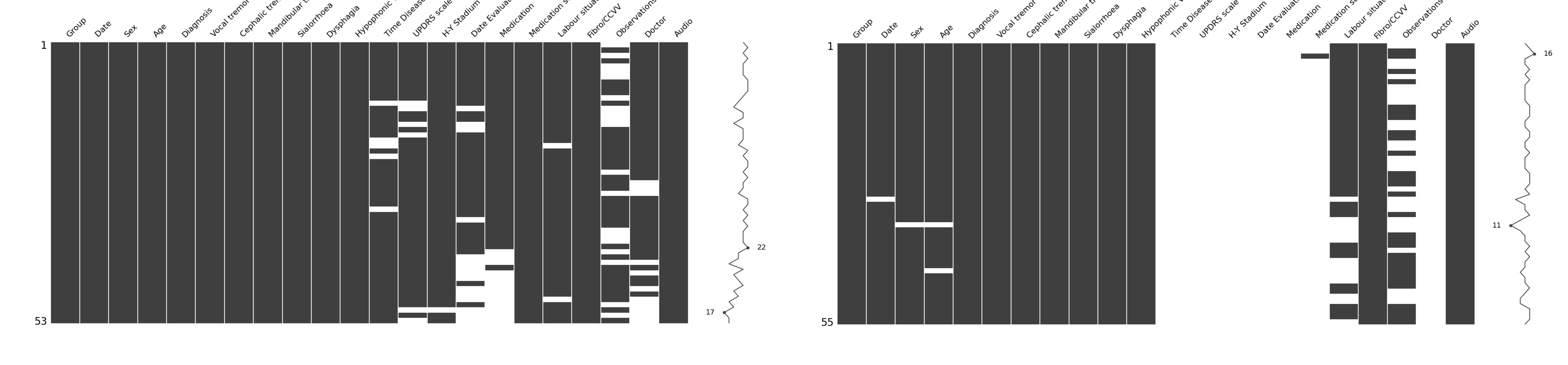}
    \caption{Percentage of missing data per column variable. Blank spaces mean no available data. Left: HC group. Right: PD group.}
    \label{fig:missigness}
\end{figure}

Technical validation was further enhanced through a naïve classification procedure to categorize the samples into \ac{PD} or \ac{HC}. Initially, features extracted using the AVCA-ByO toolbox were divided into training and testing sets, ensuring an 80-20 split by patient's ID. After normalization, these features were used to train a simple Random Forest classifier and a Logistic Regression model, yielding balanced accuracies of 64\% and 65\%, respectively.

Additionally, audio files underwent preprocessing and were classified using a pre-trained ResNet-18 \ac{CNN} from the Pytorch\textsuperscript{®} timm library. After segmenting audios into 400 ms frames with 50\% of overlap and computing mel-spectrograms, these normalized spectrograms were used to fine-tune the ResNet-18 over 10 epochs, classifying the samples into \ac{PD} or \ac{HC} with a balanced accuracy of 69\%.

These validation efforts, employing both audio features and direct audio analysis, do not aim to improve the benchmark results in the state of the art, but demonstrate the utility of the data set for the screening of \ac{PD}. Thus, the noted discrepancy between these results and those reported in previous works underscores the need for more advanced methodologies and further experiments.

\section*{Usage Notes}

\julian{The use of the dataset is restricted to a Data Usage Agreement (DUA) available in the repository \cite{neurovoz_zenodo}. Final users should copy the entire DUA into the 'request message' box at the Zenodo repository, also filling in the personal information requested. All applications that agree to the terms and conditions will be considered to access the corpus. }

The NeuroVoz dataset is complemented with a set of Python scripts for data management and analysis. These scripts were designed to provide a practical introduction to working with the dataset and to facilitate initial experiments for the screening of \ac{PD} using this corpus.

The script \texttt{"simple\_predictors\_af.py"} offers a straightforward demonstration on processing and analyzing voice quality features extracted with the AVCA-ByO toolbox, followed by a simple application using Random Forest and Logistic Regression models to classify the speech samples as either \ac{HC} or \ac{PD}. This script is intended to inspire further analytical explorations. 

For those interested in working directly with the audio files, the \texttt{"simple\_predictors\_audio.py"} script is included. This script outlines the main steps for audio preprocessing, mel-frequency spectrogram calculations, and subsequent classification with a pre-trained ResNet-18 \ac{CNN} using the Pytorch timm library. It exemplifies how to leverage \ac{DL} for \ac{PD} detection from the speech, potentially paving the way for novel research directions.

The notebook \texttt{"extract\_audio\_features.ipynb"} provides code to extract 6373 features using OpenSmile Compare 2016. This comprehensive feature extraction process is essential for detailed audio analysis and facilitates subsequent ML applications to classify speech data.

\section*{Code Availability}

The codes used for data preprocessing and cleaning can be accessed in a specific GitHub\textsuperscript{®} repository (\url{https://github.com/BYO-UPM/NeuroVoz_Dababase}). Additionally, the referenced scripts used for classification purposes are included in the same repository. 

The AVCA-ByO toolbox (\url{https://github.com/BYO-UPM/AVCA-ByO}) used to extract the voice quality features is also available as an open-source package \cite{gomez2021design}.

\section*{Acknowledgements}

This work was funded by the Ministry of Economy and Competitiveness of Spain (grants TEC-
2012-38630-C04-01, DPI2017-83405-R1, PID2021-128469OB-I00 and TED2021-131688B-I00), Ministry of Health (FIS PI21/00771), and by Comunidad de Madrid, Spain. 
\ac{UPM} also supports Julián D. Arias-Londoño through a María Zambrano UP2021-035 grant funded by European Union-NextGenerationEU.

The authors thank the Neurology and Otorhinolaryngology Services of \ac{HGUGM}, Madrid, and \ac{HUF}, Madrid, for the facilities provided, and \ac{UPM} for providing computing resources on the Magerit Supercomputer.

Finally, the authors also thank the Madrid ELLIS unit (European Laboratory for Learning \& Intelligent Systems) for its indirect support, and to all patients who selflessly participated in the study. 

\section*{Competing Interests} 
The authors declare that they have no competing interests.

\section*{Author Contributions}

J.G.G. and J.I.G. designed the experiment. J.G.G., J.D.A. and J.I.G. designed the methodology. J.M.L. and E.L.B. validated the experiment. J.M.L. and E.L.B. collected the data. J.M.L., E.L.B. and F.G.P. provided the resources. J.G.G., A.G.L. and J.I.G. wrote the initial draft version. A.G.L., J.D.A. and J.I.G. reviewed and edited the manuscript. A.G.L. provided the software to analyse the data. J.D.A. and J.I.G. supervised. J.I.G. and F.G.P. administrated the project. J.I.G. acquired the funding. All authors have read and agreed to the published version of the manuscript.

\section{Figures}

Figure 1. Sex distribution of participants.

Figure 2. Age distribution of participants.

Figure 3. Distribution of years since the diagnosis of \ac{PD}.

Figure 4. GRBAS total sum distribution of participants.

Figure 5. The free monologue is guided by asking the patient to describe the scenes presented in this illustration.

Figure 6. Percentage of completeness of each audio material for all patients. 

Figure 7. Binary indicators regarding parkinsonian symptoms.

Figure 8. Severity distribution of parkinsonian patients expressed in \ac{UPDRS} and \ac{H-Y} scales

Figure 9. Distribution of GRBAS individual evaluations for HC and PD patients

Figure 10. Database tree structure: hierarchical representation of the data records and their organization within the database

Figure 11. Percentage of missing data per column variable. Blank spaces mean no available data.

\begin{figure}[ht]
    \centering
    \includegraphics[width=\textwidth]{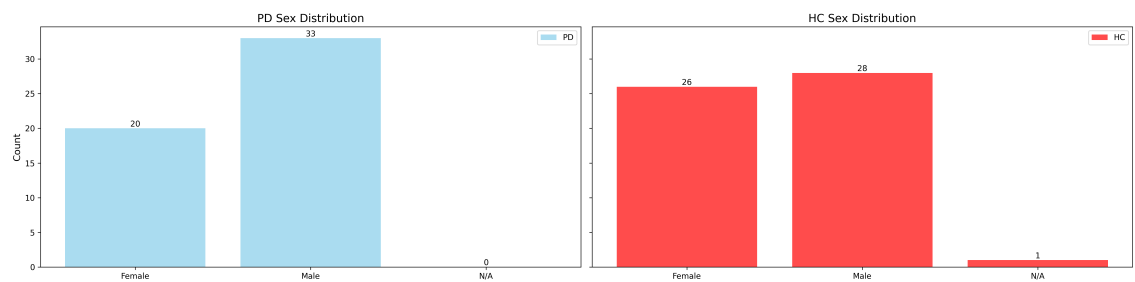}
    \caption{Sex distribution of participants. }
    \label{fig:sexdistro}
\end{figure}

\begin{figure}[ht]
    \centering
    \includegraphics[width=\textwidth]{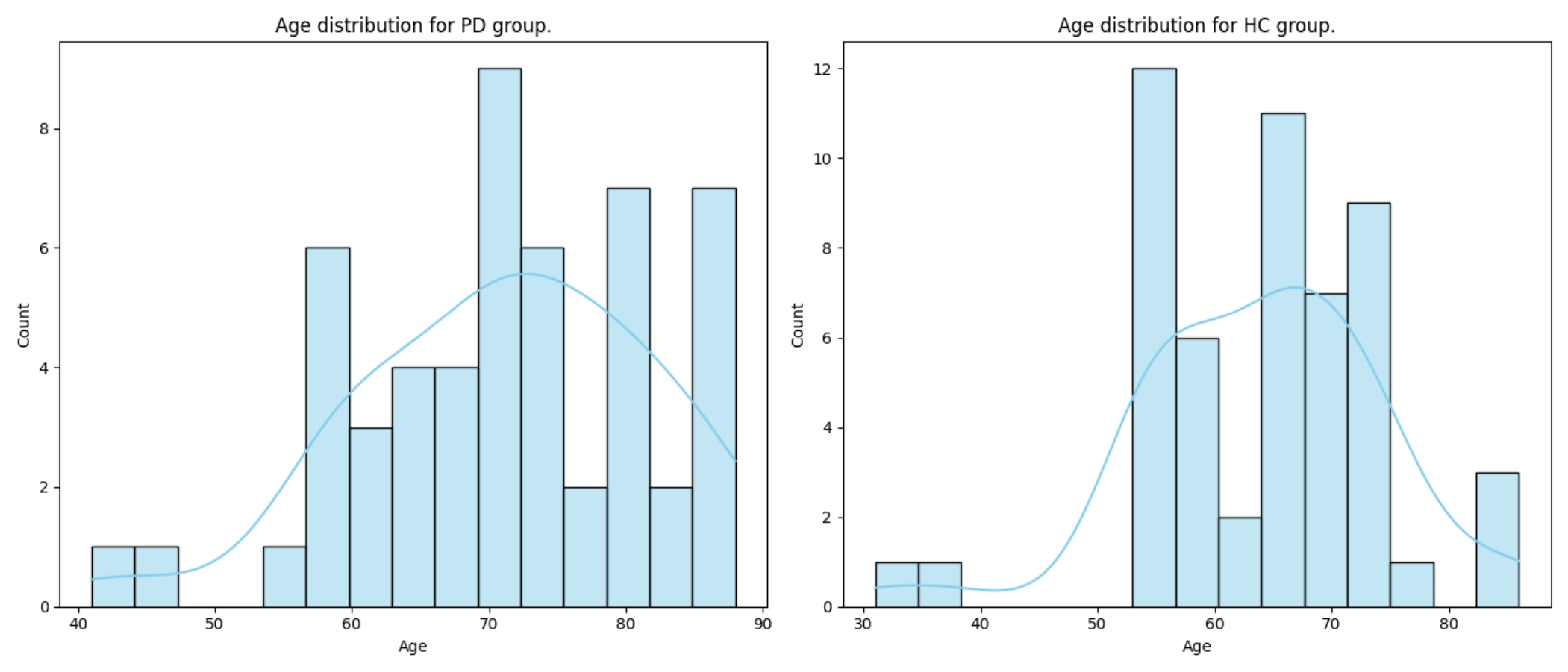}
    \caption{Age distribution of participants.}
    \label{fig:agedistro}
\end{figure}

\begin{figure}[ht]
    \centering
    \includegraphics[width=0.5\textwidth]{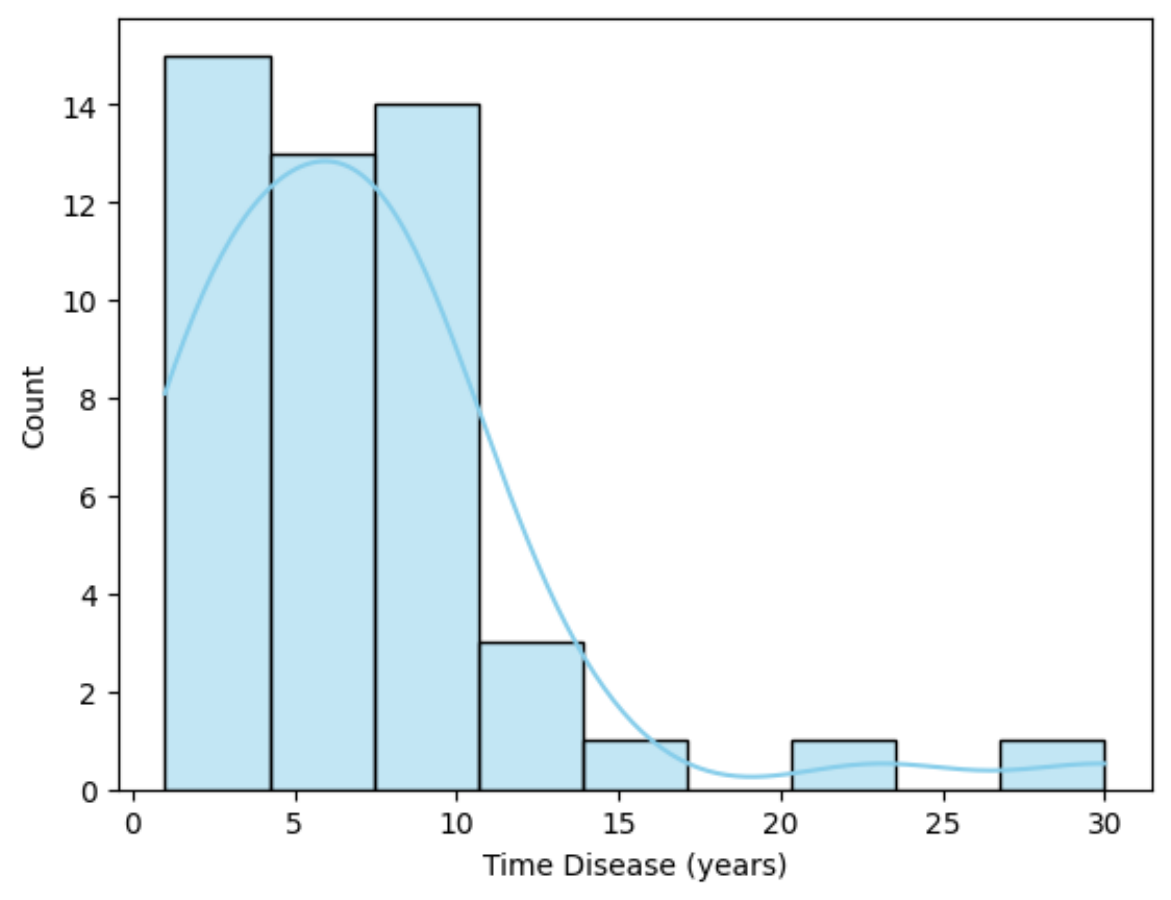}
    \caption{Distribution of years since the diagnosis of \ac{PD}.}
    \label{fig:PD_distribution}
\end{figure}

\begin{figure}[ht]
    \centering
    \includegraphics[width=\textwidth]{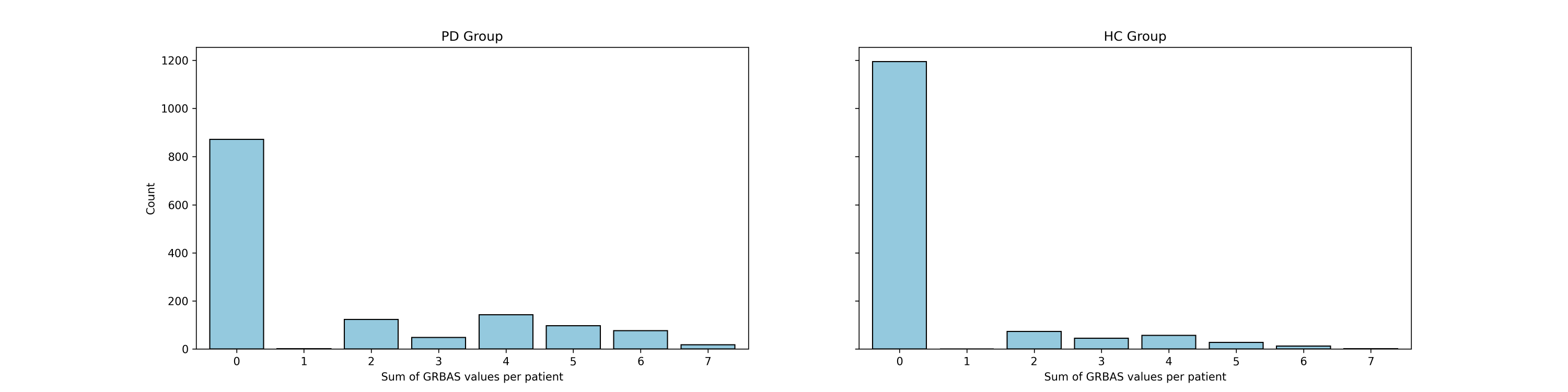}
    \caption{GRBAS total sum distribution of participants.}
    \label{fig:grbasdistro}
\end{figure}

\begin{figure}
\centering
\includegraphics[width=0.3\textwidth]{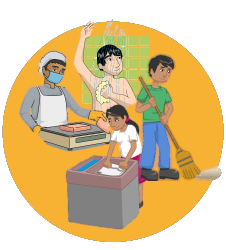}
\caption{The free monologue is guided by asking the patient to describe the scenes presented in this illustration.}
\label{fig:monologue}
\end{figure}

\begin{figure}[ht]
    \centering
    \includegraphics[width=\textwidth]{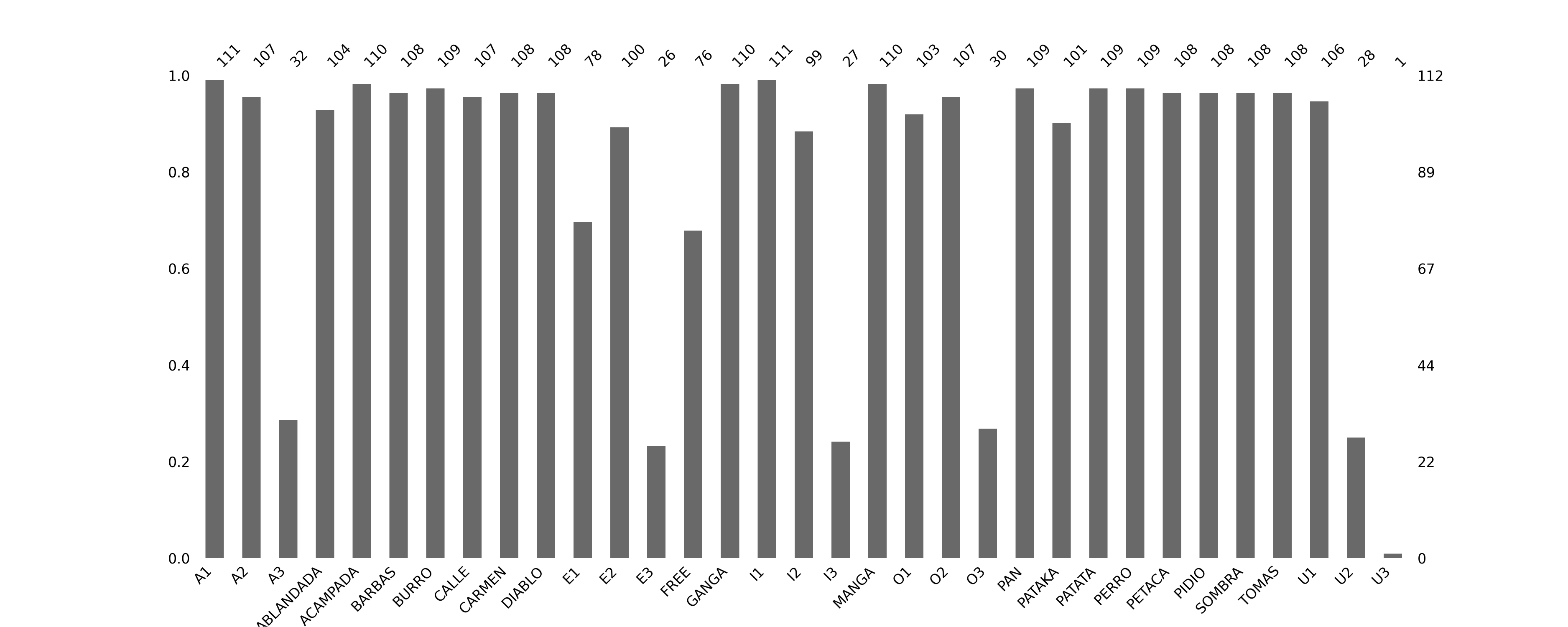}
    \caption{Percentage of completeness of each audio material for all patients.}
    \label{fig:missing_audios}
\end{figure}

\begin{figure}[ht]
 \centering \includegraphics[width=\textwidth]{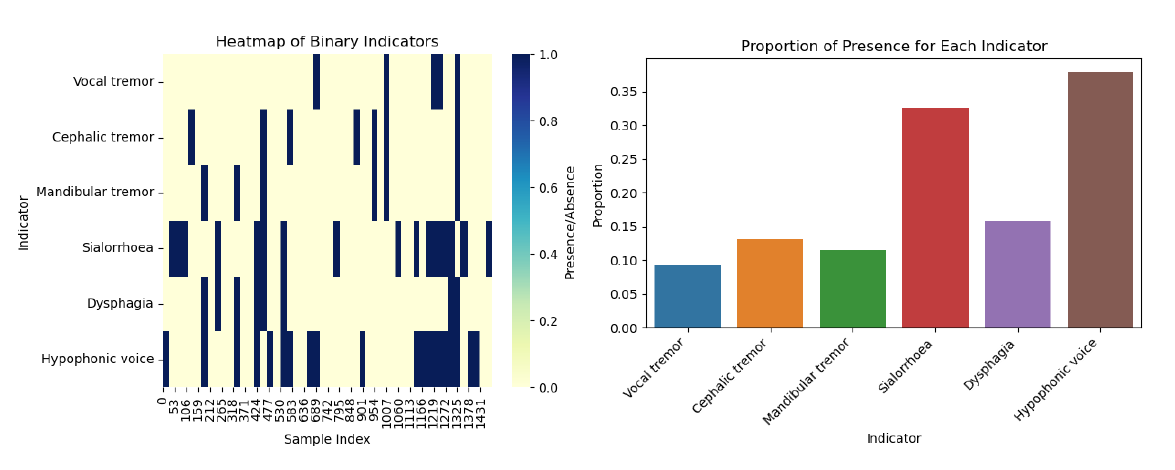}
    \caption{Binary indicators regarding parkinsonian symptoms. Left: Binary indicators of the presence of symptoms for each \ac{PD} patient. The horizontal axis determines the patient ID while the vertical axis denotes each symptom. Right: Binary proportion of presence of symptoms for each \ac{PD} patient.}
    \label{fig:binarydata}
\end{figure}

\begin{figure}[ht]
    \centering
    \includegraphics[width=\textwidth]{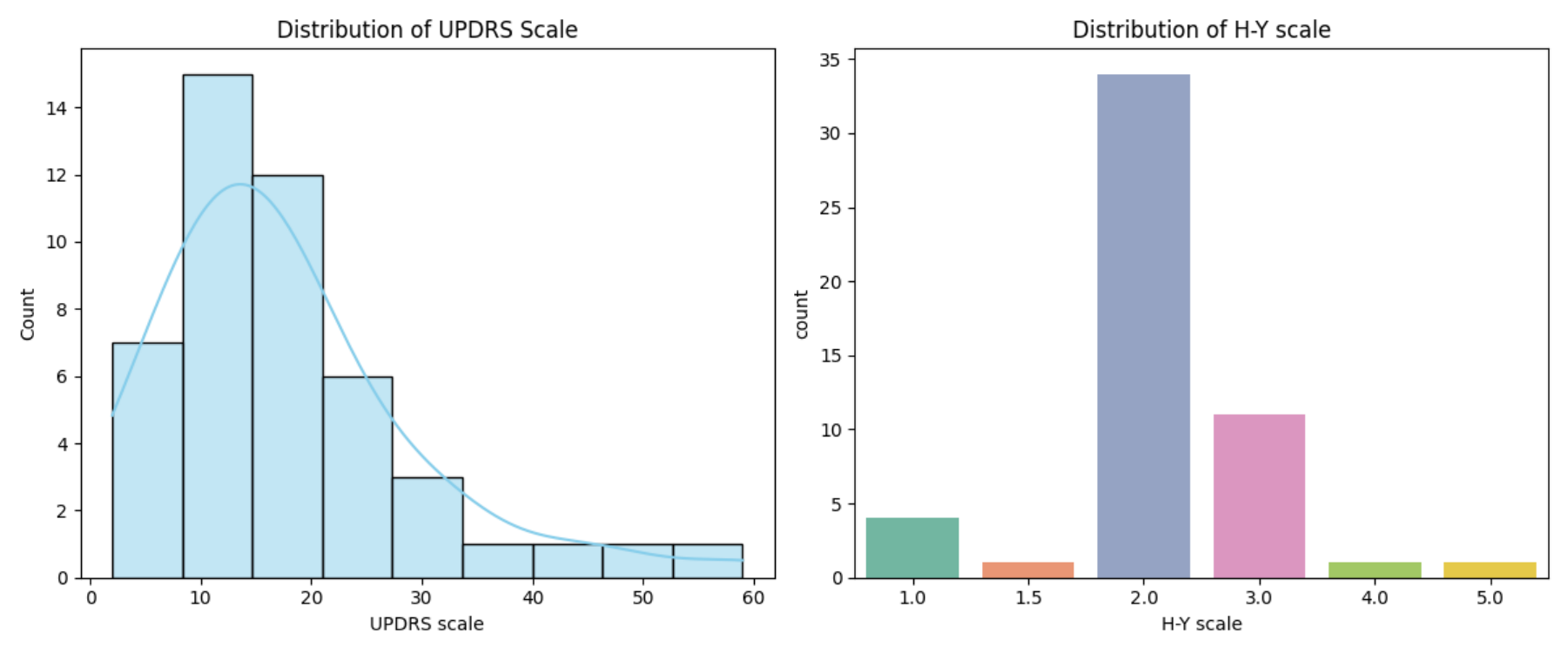}
    \caption{Severity distribution of parkinsonian patiens expressed in \ac{UPDRS} and \ac{H-Y} scales.}
    \label{fig:severitydistro}
\end{figure}

\begin{figure}
    \centering
    \includegraphics[width=\textwidth]{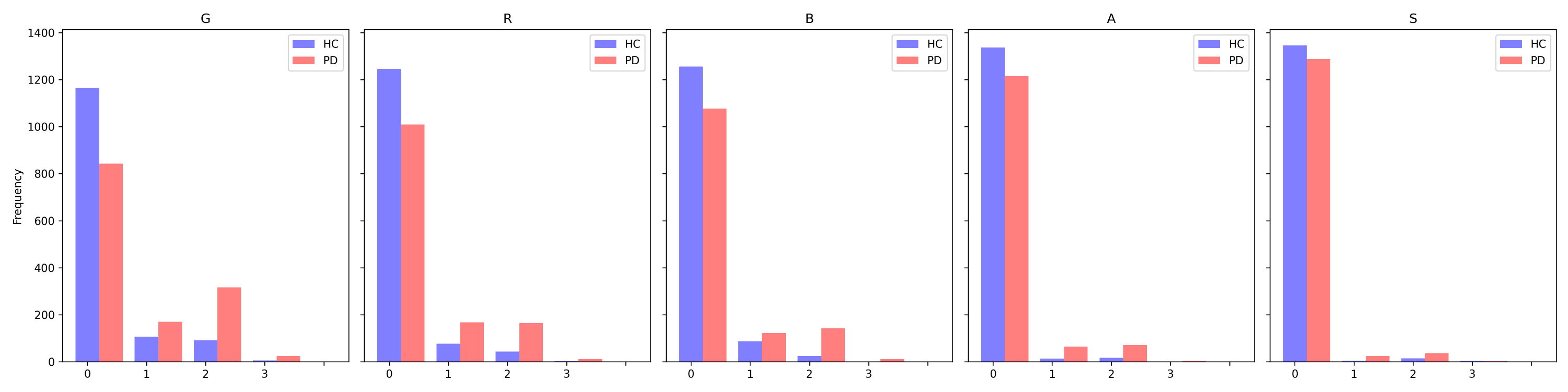}
    \caption{Distribution of GRBAS individual evaluations for HC and PD patients}
    \label{fig:grbas_dsitro}
\end{figure}

\begin{figure}
\centering
\includegraphics[width=\textwidth]{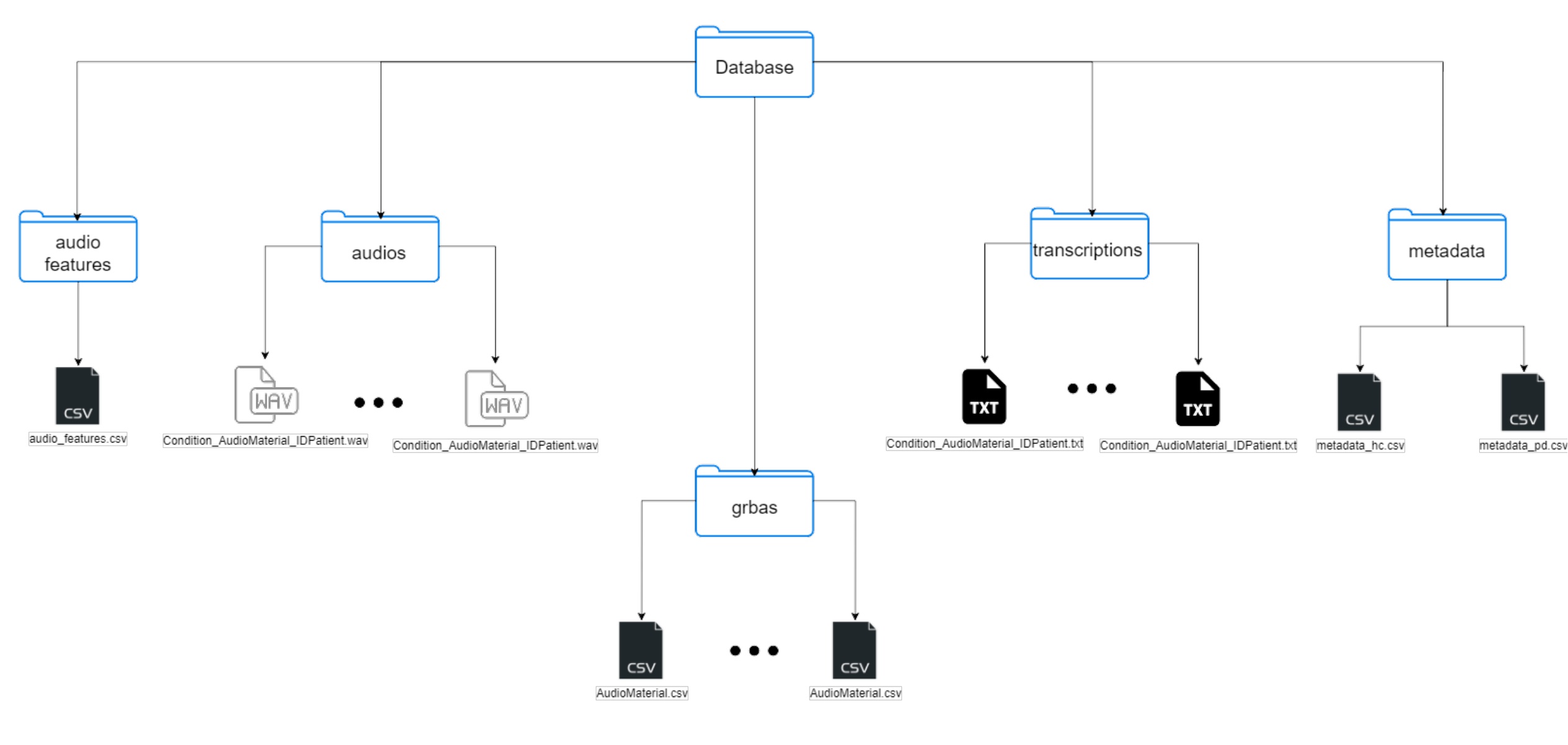}
\caption{Database tree structure: hierarchical representation of the data records and their organization within the database.}
\label{fig:db-tree-st}
\end{figure}

\begin{figure}
 \centering
    \includegraphics[width=\textwidth]{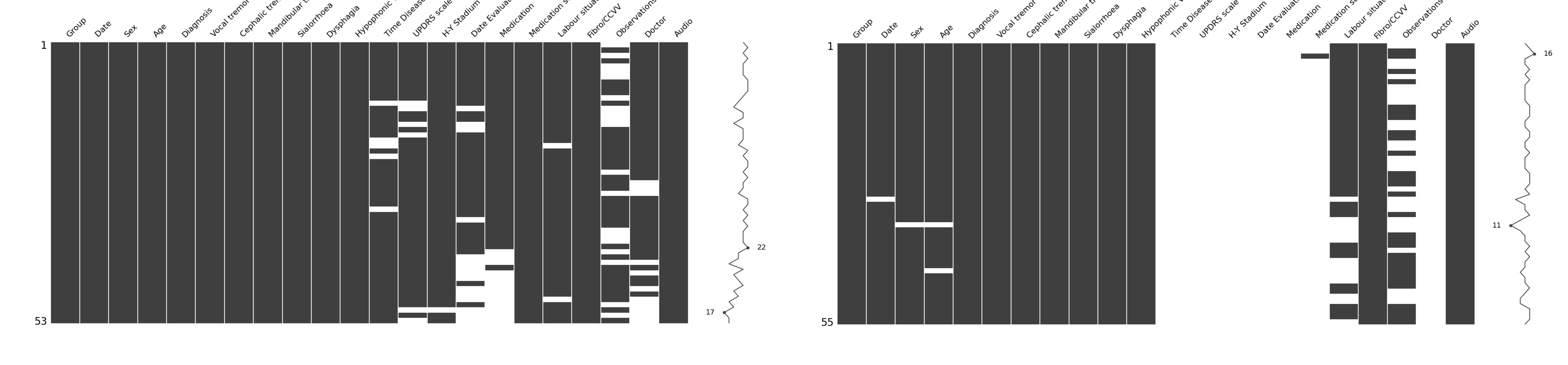}
    \caption{Percentage of missing data per column variable. Blank spaces mean no available data. Left: HC group. Right: PD group.}
    \label{fig:missigness}
\end{figure}

\section{Tables}

Table 1. Exclusion Criteria for PD and HC participants.

Table 2. Speech tasks protocol followed in order.

Table 3. Transcriptions in the \ac{IPA} and translations of selected sentences.

Table 4. Summary of the metadata collected for each participant.

Table 5. Voice measurements extracted from the recordings of the /a/ sustained vowels.

\begin{table}[h]
\centering
\caption{Exclusion Criteria for PD and HC participants.}
\label{tab:exclusion_criteria}
\begin{tabular}{p{0.45\textwidth}|p{0.45\textwidth}}
\hline
\multicolumn{2}{c}{\textbf{Common Exclusion Criteria}} \\ \hline
\multicolumn{2}{c}{A history of voice or speech alterations or laryngeal or vocal cords surgery.} \\
\multicolumn{2}{c}{Regular consumer of alcohol.} \\
\multicolumn{2}{c}{Smoker.} \\
\multicolumn{2}{c}{Speaker whose mother tongue is not Spanish.} \\
\multicolumn{2}{c}{Speaker suffering from any disease that affects speech intelligibility or affects speech or voice.} \\ \hline
\textbf{PD Patients} & \textbf{HC Participants} \\ \hline
Diagnosis of another neurological disease. & Participants with neurological disorders and/or PD. \\ \hline
\end{tabular}
\end{table}

\begin{table}[h]
\centering
\caption{Speech tasks protocol followed in order}
\label{tab:speech_task_protocol}
\begin{tabular}{p{0.1\textwidth}p{0.2\textwidth}p{0.6\textwidth}}
\hline
\textbf{Order} & \textbf{Task} & \textbf{Description} \\ \hline
1 & Phonation & Repeating sustained vowels to evaluate phonation. \\ 
2 & Velopharyngeal Closure & \ac{LR} phrases 1, 2, 6, 9, 10, 11, 12 and 13 from Table \ref{tab:IPATDU} for velopharyngeal closure assessment. \\ 
3 & Articulation & DDK test to evaluate articulation capabilities. \\ 
4 & Prosody & \ac{LR} phrases 3 and 7 from Table \ref{tab:IPATDU} for prosody evaluation. \\ 
5 & Intonation-Emotion & \ac{LR} phrases 5, 14, and 15 from Table \ref{tab:IPATDU} with emphasis on capitalised words for intonation-emotion assessment. \\ 
6 & Other tasks & \ac{LR} phrases 4, 8, and 10 from Table \ref{tab:IPATDU}. \\ \hline
\end{tabular}
\end{table}

\begin{table*}[ht]
\centering
\caption{Transcriptions in the \ac{IPA} and translations of selected sentences}
\label{tab:IPATDU}
\begin{tabular}{ p{0.5cm} p{2.3cm} p{4.3cm} p{4.3cm} p{4.3cm} }
\toprule
\textbf{Sent. \#} & \textbf{ID} & \textbf{Spanish transcription} & \textbf{IPA transcription} & \textbf{English translation} \\ \midrule
1 & ABLANDADA &  \textit{La patata no está bien ablandada} & [\textipa{la pa"tata no "esta Bjen aBlan"dada}] & ``The potato is not soft enough'' \\
2 & ACAMPADA & \textit{Mañana vamos de acampada} & [\textipa{ma"26ana "Bamos de akam"pada}] & "Tomorrow we are going camping" \\
3 & BARBAS & \textit{Cuando las barbas de tu vecino veas pelar, pon las tuyas a remojar} & [kwando las \textipa{BarBas} \textipa{de} tu \textipa{Be\texttheta ino} \textipa{Beas} pelar pon las \textipa{tu\textipa{J}as} a \textipa{remo\textipa{X}ar}] & "When your neighbor's beard you see peeling, put yours to soak" \\
4 & BURRO & \textit{Burro grande ande o no ande} & [\textipa{"bur.o "gran.de "an.de o no "an.de}] & "Big donkey, walk or not walk" \\
5 & CALLE & \textit{De la calle VENDRÁ quien de tu casa te ECHARÁ} & [de la \textipa{ka\textipa{Je}} \textipa{Bend\textipa{Ra}} \textipa{kj\textipa{en}} \textipa{de} tu \textipa{kasa} te e\textipa{tSa}\textipa{ra}] & "From outside will come who will kick you out from your house" \\
6 & CARMEN & \textit{Carmen baila el mambo} & [\textipa{"kar.men "bai.la el "mam.bo}] & "Carmen dances the mambo" \\
7 & DIABLO & \textit{Cuando el diablo no sabe qué hacer, con el rabo mata moscas} & [kwando el \textipa{\textdyoghlig}\textipa{a\textbeta lo} no \textipa{sa\textipa{Be}} ke a\textipa{\texttheta}er kon el \textipa{ra\textbeta o} mata \textipa{moskas}] & "When the devil does not know what to do, it kills flies with its tail" \\
8 & GANGA & \textit{Esto es una ganga} & [\textipa{"es.to es "u.na "gaN.ga} & "This is a bargain" \\
9 & MANGA & \textit{Juan tira de la manga} & [\textipa{"xwan "ti.ra de la "maN.ga}]  & "Juan pulls the sleeve" \\
10 & PERRO & \textit{Dame pan y llámame perro} & [\textipa{"da.me pan i ja.ma.me "pero}] & "Give me bread and call me dog" \\
11 & PAN & \textit{Al pan pan y al vino vino} & \textipa{[al pan pan i al \textbeta ino \textbeta ino]} & "To the bread, bread, and to the wine, wine
" \\
12 & PATATA & \textit{La patata blanda es buena} & \textipa{[la pa"tata "blanda es "bwena]} & "The soft potato is good" \\
13 & PETACA & \textit{La petaca blanca es mía} & [\textipa{[la pe"taka "blanka es "mi.a]}] & "The white flask is mine" \\
14 & PIDIO & \textit{No pidas a quien PIDIÓ, ni sirvas a quien SIRVIÓ} & [no \textipa{pi\textipa{Das}} a \textipa{kj\textipa{en}} \textipa{pi\textipa{\textdyoghlig}\textipa{o}} ni \textipa{si\textipa{R}\textipa{Bas}} a \textipa{kj\textipa{en}} \textipa{si\textipa{R}\textipa{B}\textipa{Jo}}] & "Do not beg the one who begged, nor serve the person who served" \\
15 & SOMBRA & \textit{El que a BUEN árbol se arrima, BUENA sombra le cobija} & [\textipa{el ke a \textbeta wen "a\textfishhookr \textbeta ol se a"rima, "bwena "somb\textfishhookr a le ko"\textbeta ixa}
] & "Who leans close to a good tree is sheltered by good shade" \\
16 & TOMAS & \textit{Tomás tira de la manta} &[\textipa{to"mas "ti.ra de la "man.ta}] & "Tomás pulls the blanket" \\ \bottomrule
\end{tabular}
\end{table*}

\begin{table}[h]
\centering
\caption{Summary of the metadata collected for each participant.}
\label{tab:metadata}
\begin{tabular}{ll}
\hline
\textbf{Data Type} & \textbf{Description} \\ \hline
Date of Recording & Date \\
Sex & Free text string \\
Age & Integer value \\
Diagnosis & Free text field \\
Symptoms & Binary variables (True/False): \\
& \quad - Vocal tremor \\
& \quad - Cephalic tremor \\
& \quad - Mandibular tremor \\
& \quad - Sialorrhea \\
& \quad - Dysphagia \\
& \quad - Hypophonic voice \\
Duration of disease & Integer value (years) \\
URPDS & \ac{UPDRS} scale \\
\ac{H-Y}  & \ac{H-Y} scales \\
Date of evaluations  & Date \\
Medication & Free text field \\
Medical treatment during recording & Annotated "ON"/"OFF" \\
Occupation & Free text string \\
Vocal folds analysis & String field (Normal/Not performed/free text) \\
Observations & Free text field for additional notes \\ \hline
Doctor & Categorical value to indicate who evaluated the speaker \\
\end{tabular}
\end{table}

\begin{table}[ht]
\centering
\caption{Voice measurements extracted from the recordings of the /a/ sustained vowels.}
\label{tab:audio-parameters}
\begin{tabular}{lcc}
\toprule
\textbf{Parameter} & \textbf{Abbreviation} & \textbf{Unit Measure} \\
\midrule
\multicolumn{3}{c}{\textbf{Perturbation Measures}} \\
\midrule
Absolute Jitter & Jitter & $\mu$Seconds \\
Relative Jitter & rJitter & \% \\
 Relative Average Perturbation & RAP & \% \\
Pitch Period Perturbation Quotient & rPPQ & \% \\
 Smoothed Pitch Period Perturbation Quotient & rSPPQ & \% \\
Absolute Shimmer & ShimmerDb & dB \\
Relative Shimmer & rShimmer & \% \\
 Amplitude Perturbation Quotient & APQ & \% \\
Smoothed Amplitude Perturbation Quotient & sAPQ & \% \\
Cepstral Peak Prominence & CPP & dB \\
\midrule
\multicolumn{3}{c}{\textbf{Noise Parameters}} \\
\midrule
 Harmonics-to-Noise Ratio & HNR & dB \\
 Cepstrum Harmonics-to-Noise Ratio & CHNR & dB \\
 Glottal to Noise Excitation Ratio & GNE & Ratio \\
 Normalised Noise Energy & NNE & dB \\
\midrule
\multicolumn{3}{c}{\textbf{Tremor Parameters}} \\
\midrule
Frequency Tremor Intensity Index & FTRI & Arbitrary Units \\
Amplitude Tremor Intensity Index & ATRI & Arbitrary Units \\
Fundamental Frequency Tremor Frequency & FFTR & Hz \\
Amplitude Tremor Frequency & ATRF & Hz \\
\bottomrule
\end{tabular}   
\end{table}

\end{document}